# An Argument for 4D Blockworld from a Geometric Interpretation of Non-relativistic Quantum Mechanics

*Michael Silberstein[1,3], Michael Cifone[3] and W.M. Stuckey[2]*

## Abstract
We use a new distinctly "geometrical" interpretation of non-relativistic quantum mechanics (NRQM) to argue for the fundamentality of the 4D blockworld ontology. Our interpretation rests on two formal results: Kaiser, Bohr & Ulfbeck and Anandan showed independently that the Heisenberg commutation relations of NRQM *follow from* the relativity of simultaneity (RoS) per the Poincaré Lie algebra, and Bohr, Ulfbeck & Mottelson showed that the density matrix for a particular NRQM experimental outcome may be obtained from the spacetime symmetry group of the experimental configuration. Together these formal results imply that contrary to accepted wisdom, NRQM, the measurement problem and so-called quantum non-locality do not provide reasons to abandon the 4D blockworld implication of RoS. After discussing the full philosophical implications of these formal results, we motivate and derive the Born rule in the context of our ontology of spacetime relations via Anandan. Finally, we apply our explanatory and descriptive methodology to a particular experimental set-up (the so-called "quantum liar experiment") and thereby show how the blockworld view is not only consistent with NRQM, not only an *implication* of our geometrical interpretation of NRQM, but it is necessary in a non-trivial way for explaining quantum interference and "non-locality" from the spacetime perspective.

[1] Department of Philosophy, Elizabethtown College, Elizabethtown, PA 17022, silbermd@etown.edu

[2] Department of Physics, Elizabethtown College, Elizabethtown, PA 17022, stuckeym@etown.edu

[3] Department of Philosophy, University of Maryland, College Park, MD 20742, cifonemc@wam.umd.edu

# 1. Introduction

We use a new distinctly "geometrical" interpretation of non-relativistic quantum mechanics (NRQM) to argue for the fundamentality of the 4D blockworld ontology. We first motivate our geometrical view (Stuckey *et. al.* 2006) by distinguishing between "principle" and "constructive" approaches to quantum theory and spacetime theory. By taking a principle approach to both, we argue for a geometrical interpretation whose fundamental ontology is one of spacetime relations as opposed to constructive entities whose time-dependent behavior is governed by dynamical laws. Our view rests on two formal results: Kaiser (1981 & 1990), Bohr & Ulfbeck (1995) and Anandan, (2003) showed independently that the Heisenberg commutation relations of NRQM *follow from* the relativity of simultaneity (RoS) per the Poincaré Lie algebra, and Bohr, Ulfbeck & Mottelson (2004*a* & 2004*b*) showed that the density matrix for a particular NRQM experimental outcome may be obtained from the spacetime symmetry group of the experimental configuration. Together the formal results imply that contrary to accepted wisdom, NRQM, the measurement problem and so-called quantum non-locality do not provide reasons to abandon the 4D blockworld implication of RoS. But rather, the deep non-commutative structure of the quantum and the structure of spacetime as given by the Minkowski interpretation of special relativity (STR) are unified in a 4D spacetime regime that lies between Galilean spacetime (G4) and Minkowski spacetime ($\mathcal{M}^4$). After discussing the full philosophical implications of these formal results, we motivate and derive the Born rule in the context of our ontology of spacetime relations by appropriating Anandan (2002). Taken together the aforementioned formal results allow us to model NRQM phenomena such as interference without the need for realism about 3N Hilbert space, establishing that the world is really 4D and that configuration space is nothing more than a calculational device. Our new geometrical interpretation of NRQM provides a geometric account of quantum entanglement and so-called non-locality free of conflict with STR and free of interpretative mystery. We also provide a novel statistical interpretation of the wavefunction that deflates the measurement problem. Finally, we apply our explanatory and descriptive methodology to a particular experimental set-up (the so-called "quantum liar experiment") and thereby show how the blockworld view is not only consistent with NRQM, not only an *implication* of our geometrical interpretation of NRQM, but it is necessary in a non-trivial way for explaining quantum interference and "non-locality" from the spacetime perspective.

In section 2 we argue that STR unadorned implies a blockworld (the reality of all events, past, present and future). In section 3 we provide the necessary historical and conceptual background to appreciate the complex relationship between the formal structures and various interpretations of both NRQM and STR. This allows us to motivate our principle geometric account of both theories. Section 4 is devoted to an explication of the Kaiser *et. al.* results and their philosophical implications. Likewise, the Bohr *et. al.* results and their implications are the subject of section 5. Our fundamental ontology of spacetime relations is formalized and the Born rule derived on that basis in section 6. As well, our geometric interpretation of quantum entanglement and "non-locality" is presented, as is our statistical interpretation of the wavefunction. In section 7, for purposes of demonstration, we bring to bear the full methodological and metaphysical apparatus of our geometric interpretation upon the "quantum liar experiment." We conclude with some final thoughts in section 8.

## 2. Special Relativity, Relativity of Simultaneity and the Blockworld

Given that this is well-known and well-worn territory we will be brief in our demonstration that given the relativity of simultaneity, special relativity unadorned implies a blockworld[1]. We realize that not everyone grants this point, but we leave those arguments for the other chapters in this volume. And there are those who grant this point but argue that STR must be so adorned given NRQM, general relativity or some version of quantum gravity. In this paper we shall respond only to the argument from NRQM against the BW[2]. Consider the following example adopted with minor modifications from DeWitt and Mermin[3,4].

Our adaptation to the DeWitt/Mermin version of this example is to consider only local observations. We do this to emphasize that the BW implication of RoS is not an optical illusion resulting from the finite speed of light. In order to keep our observations local we add three new characters – Bob, Alice and Kim – to Joe and Sara of the DeWitt/Mermin version. The boys, Bob and Joe, are at rest with respect to each other and the girls are at rest with respect to each other. Joe and Bob see the girls moving in the positive x direction at 0.6c (Figure 1). The girls, therefore, see the boys moving in the negative x direction at 0.6c (Figure 4). Who is *actually* moving?

The answer to this question is central to the BW perspective. According to the first postulate of STR, there is no way to discern absolute, uniform motion so *either perspective is equally valid*. The girls are correct in saying it is the boys who are moving, and the boys are correct in saying it is the girls who are moving. This is equivalent to saying there is no preferred, inertial frame of reference in the spacetime of STR. Now we compare some observations and their consequences.

Joe is located at x = 0 (lower case coordinates are the boys') and Bob is at x = 1000km. Joe says Sara's clock read T = 0 as she passed him (upper case coordinates are the girls'). Joe's clock likewise read t = 0 at that event. Bob said Kim's clock read T = -0.0025s when she passed him. Bob's clock read t = 0 at that event (Figure 1). The girls agree with these mutual clock readings, so what's the problem?

The boys synchronized their clocks so the events Joe/Sara (event 1) and Bob/Kim (event 2) are simultaneous, having both occurred at t = 0. Clearly, according to the boys, the girls' clocks are not synchronized. Unfortunately, the girls also synchronized their clocks so, according to the girls, events 1 and 2 are *not* simultaneous. Who's right?

Neither of their frames is preferred, so the girls *and* the boys are correct! Whether or not space-like separated events (Figure 2) are simultaneous is relative to the frame of reference. So, what is the consequence of this "relativity of simultaneity?"

RoS renders the view known as "presentism" highly suspect. Presentism is the belief that everyone shares a unique, 'real' present state while their common past states no longer exist and their common future states are yet to exist. According to presentism, everyone could agree with a statement such as, "Sam is surfing in California while Jonathan is shoveling snow in New York." If Sam is 25 years old, the 24 year-old version of Sam no longer exists, i.e., is no longer 'real', and the 26 year-old version will not exist, i.e., will not be 'real', for another year. There is a sense that we share in a 'real' present moment with everything else in the universe and this attribute of 'realness' moves along all worldlines synchronously (Figure 3). In fact, what one means by "the universe" is vague unless everyone agrees on a spatial surface of simultaneity in spacetime. But, as we continue with our example, it should become clear that RoS is contrary to this commonsense notion of presentism. Let us continue.

---

[1] BW for short.

[2] *See* Saunders (2002) and Petkov (2005 and 2006) for three thorough refutations of most of the counter-claims mentioned above.

[3] DeWitt (2004), 213 – 219. DeWitt says his example "owes much to" Mermin (1968).

[4] Calculations are found in the Appendix.

Since Kim's clock read T = -0.0025s at event 2, the girls say Bob passed Kim 0.0025s *before* Joe got to Sara at T = 0 (Figure 4). The event simultaneous with event 1 is Bob passing Alice at T = 0, i.e., event 3 (Figure 5). All agree that Alice's clock read T = 0 when Bob was there, but they also agree that Bob's clock read t = 0.002s when he was at Alice's position. Here's how the girls tell the story.

Sara is at X = 0, Alice is at X = 800km and Kim is at X = 1250km. The boys are not 1000km apart, as they claim, but rather only 800km apart. The girls know this since Bob was at Alice's position (X = 800km, T = 0) when Joe was at Sara's position (X = 0, T = 0). As a consequence, Bob passed Kim (event 2) *before* Joe got to Sara (event 1). In fact, it took Bob 0.0025s to get from X = 1250km to X = 800km moving at 0.6c, that's why Kim's clock read -0.0025s when Bob was there. So, who shares the attribute of 'realness', i.e., where is the spatial frame of 'realness' which defines "the universe?"

Well, unless we can touch things which are not 'real', Joe and Sara are 'co-real' at event 1, and Bob and Kim are 'co-real' at event 2 (Figure 6). But, Joe and Bob are 'co-real' at events 1 & 2 (t = 0 for both) so we see that all four characters involved in events 1 & 2 are 'co-real'. This means Sara shares 'realness' at T = 0 with Kim at T = -0.0025s, *and* Sara shares 'realness' at T = 0 with Kim at T = 0. Thus, Kim at T = 0 shares 'realness' with Kim at T = -0.0025s so *RoS implies the past is as 'real' as the present*. Now let's look at the boys' perspective.

Joe is at x = 0 and Bob is at x = 1000km. Sara and Kim passed the boys (events 1 & 2) at t = 0, so Joe and Bob do not believe Sara and Kim are 1250km apart, but only 1000km apart. Alice passed Bob 0.002s later, so she must be 0.6c x 0.002s = 360km behind Kim (not 450km as the girls claim). At event 3, Alice shares 'realness' with Bob while she shares 'realness' with Sara (both T = 0, Figure 6). However, at T = 0 Sara also shares 'realness' with Joe (both at event 1). Thus, Bob at t = 0 shares 'realness' with Bob at t = 0.002s so *RoS implies the future is as 'real' as the present*. And together, these conclusions imply a blockworld, by which is meant that *all of the past, present and future are equally real.*

## 3. NRQM, Relativity of Simultaneity and Blockworld: Orthodoxy & Heterodoxy

Quantum theory[5] *simpliciter* and special relativity are not in conflict, but rather it is only on some *interpretations* of quantum theory where conflict arises. For example, the following interpretations of quantum theory are consistent with relativity, requiring no preferred frame: Tumulka-GRW (Maudlin 2005 ms.); hyperplane-dependent collapse

---

[5] We adopt the convention that quantum *theory* refers to a certain abstract and very general structure, whereas quantum *mechanics* refers to a particular instantiation of that structure with an "interpretation." Interpretations, in general, supply quantum theory with a *physical ontology* (and perhaps supplemental dynamical laws) with which to model the world in terms of the theory.

accounts (Maudlin 1998); Saunders-Wallace-Everett (Brown & Timpson 2002) to name a few.

Even though there is no *necessary* tension between the basic structure of quantum theory (the structure of physical states allowed by Hilbert space and how those states evolve over time) and the structure of STR (the structure of spacetime events given by the "causal" or Minkowski spacetime geometry necessitated by the two postulates of STR), there is a question as to how to understand the relationship between the theory of the quantum and the theory of relativistic spacetime structure (special relativity). This is a question about how to interpret the *structures of the theories themselves*.

Increasingly in the literature a divide is forming between interpretations of special relativity and quantum theory along the so-called "constructive" vs. "principle" axis of theory interpretation. While some might question whether, in practice, such a distinction is useful, or what the metaphysical/epistemological import of such a distinction is, many find it a useful conceptual device in itself. Roughly, the distinction amounts to the distinction one can draw between, say, the axioms of geometry and a model or instantiation of those axioms. In general, a "principle" theory in the natural sciences is one where a set of axioms – or *physical* postulates[6] – are outlined, that entail a characteristic structure which our universe may exhibit. A "constructive" theory is usually associated with some principle theory, of which it is a particular instantiation, although not necessarily so. Constructive theories supply some physical ontology (e.g., Newtonian point-particles for statistical mechanical theories) and a dynamics (again, Newtown's laws of motion) which are supposed to "underwrite" the merely phenomenological laws of some other theory (in this case, thermodynamics – laws that refer to "gasses" or "heat" or "entropy," etc.). The "underwriting" has its cash-value in the ability to re-derive or re-state the *essential content* of the principles, but in more concrete, visualizable terms[7] (and perhaps in a way that allows the theorist to easily derive *predictions* from the more abstract principles of the theory, principles which might otherwise have no obvious reference to the experiences of scientists in their laboratory).

The orthodox view is that STR as it now stands is a principle theory and most interpretations of NRQM are constructive accounts because most people assume that the theory is about quantum constructive entities and the dynamical laws that govern them. However, as we will see shortly, there is some disagreement as to whether STR requires a constructive interpretation in order to be complete, whether or not NRQM is that constructive theory grounding STR, and there is even some disagreement as to whether NRQM is best viewed as a constructive theory at all.

But as to what that structure, given by the principles, refers, this will depend on the *kind* of theory under consideration. For spacetime theories, that structure is the metrical structure of *spacetime events* ("happenings" at particular times and places). It is a good bit trickier for quantum theories, since it is by no means clear how to relate them to other more familiar theories like classical theories (or classical *mechanical* theories). And it is tricky to even say exactly what the principles refer to in the world – "measurement acts," the behavior of "matter," "information," etc. Furthermore, and perhaps more troubling, not everyone agrees on what the postulates are! Does quantum theory include or exclude the collapse postulate, for example (as von Neumann's famous axiomatic presentation seems to take for granted)? Another point of disagreement arises here, namely, on the proper relationship between the so-called spacetime background (the geometry of the world) and the constructive dynamical entities "embedded" in that background.

---

[6] Some have thought that even the postulates of mathematical geometry are "physical" in some sense, which would make even geometry a kind of "natural science."

[7] Indeed, Hertz thought that part of the function of such theories was to provide for a picture or some kind of literal representation of the world given by the theory (Frisch 2005).

Nevertheless, as many in the quantum-logical camp were eager to point out[8], everyone can agree quite easily on a couple of basic structural features which are essential to quantum theory: a non-Boolean lattice structure of *measurement propositions*. Such a logical structure will capture such quantum-theoretic features like "interference" or "uncertainty" as a characteristic structure of what can and cannot be *simultaneously measured* according to the theory (and we can indeed represent classical mechanical theories like this too, conveniently)[9]. Also, one can, with this same logical structure, represent the characteristic *structure of correlations* that quantum theory (and any of its interpretations) exhibits (that structure being the well-known Bell correlations).

A constructive interpretation of *this* structure – the non-Booleanity of measurement propositions and the structure of entangled quantum states – would amount to providing some kind of physical ontology (particles, fields, wavefunctions) and a dynamics of how that structure *changes over time* in accordance with the essential features of quantum theory. This ontology-plus-dynamics would also have to reproduce the characteristic structure of correlations for non-locally entangled quantum mechanical systems. This is the fundamental challenge to natural philosophers today, aside from how to relate the theory to Minkowski spacetime.

In relativity theory, we have two physical postulates (relativity and light postulates) and we have a *geometric model* or "interpretation" of those postulates – Minkowski's hyperbolic 4-geometry that gives us a geometry of "light-cones." The "blockworld" view, as demonstrated in section 2 with a simple physical example, tries to establish a *metaphysical* interpretation of the Minkowski geometrical rendition of special relativity. It is a view that tries to establish the reality of all spacetime events, whose structure is given by the special relativistic metric. It does *not* try to find an ontology *per se*. This would amount to defending a view about how "spacetime events" relate to the objects of our experience (like cars, tables, falling empires, and swirling galaxies)[10]. In this sense, therefore, the blockworld view of Minkowski spacetime does *not* commit one to an ontology *of* those spacetime events – just their equal reality. So the Minkowski interpretation of the postulates of relativity do not constitute a "constructive theory" of spacetime. Needless to say, the blockworld does not commit one to either a constructive *or* principle interpretation of special relativity. Though, since the blockworld is a metaphysical interpretation of the geometrical model of special relativity, and since such a model *does not add an ontology per se*, the blockworld view is more naturally associated with a principle account of STR.

---

[8] Though, even as Putnam (2005) has recently pointed out, the quantum-logical school of interpretation really does not resolve, so much as clarify the logical structure of, the fundamental interpretive problems with quantum theory.

[9] That is, aside from providing a dynamics of "beables" with which to reconstruct quantum theory *constructively*, one can (in an *interpretively neutral* way) provide the structure of *observables*, whose reference is to acts of measurement on physical systems.

[10] E.g., everywhere-continuous spacetime "worms" (the 4D view), or infinitely thin slices of space (the 3+1 view) with some additional affine connection liking each slice to the next.

*3.1 Special Relativity and Quantum Theory.* Most natural philosophers are inclined to accept that special relativity unadorned implies the blockworld view. Among those who might agree that special relativity unadorned implies a blockworld are those who think that quantum theory provides an excellent reason to so adorn it. That is, there are those who claim that NRQM non-locality or some particular solution to the measurement problem (such as collapse accounts) require the addition of, or imply the existence of, some variety of preferred frame (a preferred foliation of spacetime into space and time)[11]. This trick could be done in a number of ways and *need not* involve postulating something like the "luminiferous aether." For example, one could adopt the Newtonian or neo-Newtonian spacetime of Lorentz[12], or one could *add* a physically preferred foliation to $\mathcal{M}^4$.

Most of these moves, however, lack an answer to a deeper question: how *exactly* are special relativity and quantum theory related? In other words, to pursue our earlier analogy, is quantum theory – or some interpretation of it – the "statistical mechanics" that underwrites the "thermodynamics" which is special relativity? Moreover, what is the right ontology of quantum theory, and how does one avert the standard litany of conceptual problems with that ontological interpretation? If the conceptual function of a theory like statistical mechanics is to provide a physical ontology *from which one can reconstruct the laws of the macroscopic phenomena from the underwriting laws of the (constitutive) microscopic phenomena,* then most of the attempts to argue for a "preferred foliation" on the basis of quantum theory fall rather short. None of these moves that invoke quantum theory are intended to provide something like a Lorentzian underpinning to the postulates of Einstein's special relativity *on the basis of quantum theory as the right account of the behavior of matter.* Most simply argue something like, for example, "*if* there is collapse, *then* some spatial hypersurface must be physically preferred." Such a natural philosopher will then try to establish the truth of the antecedent, but its consequent is *merely an existence claim.* What exactly *constitutes* that physical frame in spacetime? Is it itself "made out of" quantum-mechanical constituents? These types of arguments are indeed damaging to blockworld (if the antecedent can be established), but too quick in the final analysis (since it is by no means clear that the consequent is defensible on quantum-theoretical grounds).

*3.2 Quantum Theory underneath Special Relativity?* There is, however, one notable exception to this lopsidedness: Harvey Brown has tried to defend the *heterodoxical* view that special relativity *requires* a constructive, underwriting *theory of matter* from which one can recover (at least in principle) the phenomenological postulates of special relativity. This move requires defending two claims: (1) special relativity can be given an empirically equivalent constructive interpretation *without the necessity of re-introducing a preferred frame from the outset* (2) quantum theory can be invoked as the long-awaited theory of matter upon which one can reconstruct the postulates of relativity (without thereby denying the truth of those postulates in the process[13]). Brown defends (1) rather thoroughly, but leaves (2) somewhat vaguely defended. Let us characterize this heterodox view in more detail.

---

[11] *See* Tooley (1997) ch. 11, for but one example.
[12] As will be discussed shortly, Brown (2005) develops a sophisticated neo-Lorentzian account of spacetime structure from a "dynamical perspective."

As we said, the orthodox view of STR, as Einstein conceived it, is that it is a principle theory about kinematics and Minkowski provided a unified geometric interpretation of the principles where space and time form some kind of whole. However, as many have pointed out recently[14], Einstein's principle approach to the problem of devising an adequate "electrodynamics of moving bodies[15]" was a move he made out of "desperation." All other things being equal, a constructive theory is to be preferred, which provides for, as Lorentz and Hertz might put it, "true physical insight" (Frisch 2005). So, in light of this preference (and assuming that only constructive theories provide "true physical insight[16]"), special relativity's ultimate constructive or "underwriting" story *has been left an open question*, one to be filled in by our best theory of matter. Presently, so this view goes, that is the quantum theory. Therefore, as it stands now, quantum theory is the constructive theory of matter that will *complete* the principle theory of space and time Einstein found. This is the heterodox view, an argument for which Brown attempts to articulate and defend in great detail with his recent book *Physical Relativity: Spacetime Structure from a Dynamical Perspective* (2005).

Since it is taken largely for granted that quantum theory is a theory of the fundamental structure and nature of *matter*[17], such a theory *could* be the long-awaited *constructive* theory Einstein despaired over with his principle version of STR, and that Lorentz desired but ultimately failed to find. In order to tell this story, however, Brown defends – quite in contrast to the received view – a sophisticated *constructive* account of STR, whose aim is to ultimately defend a "dynamical" account of spacetime structure:

> in a nutshell, the idea is to deny that the distinction Einstein made in his 1905 paper between the kinematical and dynamical parts of the discussion is a fundamental one, and to assert that relativistic phenomena like length contraction and time dilation are in the last analysis the result of structural properties of the quantum theory of matter (2005, vii-viii).

---

[13] It is important to point out that for this move to be well-motivated, it ought to be *at least possible* for one to take quantum theory as the *fundamental* and/or *universal* theory of matter *without* thereby impugning either postulate of relativity. For example, since Bohmian mechanics *does* (quite radically) violate Lorentz invariance *at the level of the beables* (i.e., the underwriting physical ontology), such an interpretation of quantum theory is suspect as the "underwriting" theory of special relativity (since it denies the truth of STR at a *fundamental level!*). Since Brown has recently ended support for Bohmian mechanics (Brown & Wallace, 2005), and has explicitly argued that Everett does not *prima facie* conflict with special relativity (or *any* theory of space and time, for that matter; *see* Brown & Timpson 2002), it seems plausible that Brown would endorse an Everett-style interpretation of quantum theory.

[14] Frisch 2005; *see also* Janssen 2002*b*.

[15] The title of Einstein's famous 1905 paper.

[16] A premise which we *reject* quite explicitly, though on the basis of our radical relational ontology. *See* sections 5 and 6 of this paper for how our relationalism is justified and successfully implemented (i.e., in the derivation of the Born rule).

[17] A claim disputed by many who argue for quantum theory as a kind of information theory (e.g., Bub 2004 and 2005; *see* Hagar 2003, Hagar and Hemmo 2006 and Brown and Timpson 2006 for some criticisms). For these philosophers, the question of the structure of matter (or its inner constitution) is largely beyond the scope of quantum theory itself, whose principles are about the *structure of information* that can be *communicated* between physical systems – irrespective of their constitution.

With a constructive theory of STR in hand, perhaps along Brown's line, one might attempt to block the blockworld interpretation. As Callender notes (2006, 3):

> In my opinion, by far the best way for the tenser to respond to Putnam *et. al.* is to adopt the Lorentz 1915 interpretation of time dilation and Fitzgerald contraction. Lorentz attributed these effects (and hence the famous null results regarding an aether) to the Lorentz invariance of the dynamical laws governing matter and radiation, not to spacetime structure. On this view, Lorentz invariance is not a spacetime symmetry but a dynamical symmetry, and the special relativistic effects of dilation and contraction are not purely kinematical. The background spacetime is Newtonian or neo-Newtonian, not Minkowskian. Both Newtonian and neo-Newtonian spacetime include a global absolute simultaneity among their invariant structures (with Newtonian spacetime singling out one of neo-Newtonian spacetime's many preferred inertial frames as the rest frame). On this picture, there is no relativity of simultaneity and spacetime is uniquely decomposable into space and time. Nonetheless, because matter and radiation transform between different frames via the Lorentz transformations, the theory is empirically adequate. Putnam's argument has no purchase here because Lorentz invariance has no repercussions for the structure of space and time. Moreover, the theory shouldn't be viewed as a desperate attempt to save absolute simultaneity in the face of the phenomena, but it should rather be viewed as a natural extension of the well-known Lorentz invariance of the free Maxwell equations. The reason why some tensers have sought all manner of strange replacements for special relativity when this comparatively elegant theory exists is baffling.

*3.3 The Heterodoxy of our Geometric Interpretation.* Part of this paper is an extended reply to both the orthodox view, and the new heterodoxy. Whereas most orthodox interpreters of special relativity, when trying to defeat the blockworld view, use quantum theory simply to establish the existence of a preferred frame *without* answering the deeper question as to how exactly *and ontologically* the spacetime structure of relativity is related to quantum theory (or one of its many interpretations), we provide an answer to that question with a radically new geometric interpretation of NRQM. Such an ontology, as we show, provides for not only a rather natural transition from classical to quantum mechanics, but also resolves – deeper down and *prior to* considerations about relativistic invariance, etc. – the conceptual tensions endemic to quantum theory in a relativistic context. It is here that we also reply to the orthodoxy, which holds that quantum theory is a theory of the behavior of matter-in-motion and Brown's heterodox view (though perhaps soon to be orthodoxy) that special relativity stands in need of constructive theoretical completion by quantum theory. Our view is that quantum theory can be interpreted as a theory of principle, but one which provides a further structural *constraint* on the introduction of events in *spacetime*, and that quantum *phenomena* can be modeled in spacetime *without the necessary invocation of or realism regarding Hilbert space geometries*. In this way, our heterodoxical view marries a principle interpretation of special relativity with *a principle interpretation of quantum theory*. This is the heart of *our* heterodoxy.

We take this to be a radically new heterodoxy not only because of our irreducibly principle interpretation of both STR and quantum theory, but also because our ontology *collapses the matter-geometry dualism with an ontology of spacetime relations*. As will be detailed in what follows, our interpretation of both STR and NRQM is a brand of ontological structuralism. Furthermore, our view defends the surprising thesis that the relativity of simultaneity plays an *essential role* in the spacetime regime for which one can obtain the Heisenberg commutation relations of non-relativistic quantum mechanics – the cornerstone of the structure of quantum theory.

This point bears repeating. While it is widely appreciated that special relativity and quantum theory are not necessarily incompatible, what is *not* widely appreciated are a collection of formal results showing that quantum theory and the relativity of simultaneity are not only compatible, but in fact are *intimately related*. More specifically, in the present paper we will draw on these results and clearly show that it is precisely this "nonabsolute nature of simultaneity[18]" which survives the $c \rightarrow \infty$ limit of the Poincaré group, and which *entails* the canonical commutation relations of *non-relativistic* quantum mechanics. These results lead us to formulate a new *geometric* account of NRQM that will be elucidated in later sections of the paper.

We will also show that this geometric interpretation of NRQM nicely resolves the standard conceptual problems with the theory: (i) *prior to* the invocation of any interpretation of quantum theory itself and (ii) *prior to* the issue of whether any interpretation of quantum theory – i.e., a *mechanics* of the quantum – can be rendered relativistically invariant/covariant. Namely, we will provide both a geometrical account of entanglement and so-called "non-locality" free of tribulations, *and* a novel version of the statistical interpretation that deflates the measurement problem. Our geometrical NRQM has the further advantage that it does not lead to the aforementioned problems that some *constructive* accounts of NRQM face when relativity is brought into the picture, such as Bohmian mechanics and collapse accounts like the wavefunction interpretation of GRW. On the contrary, not only does our view require no preferred foliation but it also provides for a profound, though little-appreciated, *unity* between STR and NRQM *by way of the relativity of simultaneity*[19].

*3.4 Our geometrical interpretation in a nutshell.* To summarize, our view can be characterized as follows:

(i)     We are realists about the geometry of spacetime but antirealists about Hilbert space.

(ii)    We adopt the view that NRQM is a principle, not a constructive, theory in the following respects:

a.   it merely provides a probabilistic rule by which new trajectories are generated – i.e., we take NRQM *qua* principle theory to provide *constraints on the introduction of events in spacetime*.

---

[18] Kaiser (1981), p. 706.

[19] In this respect, our interpretation is close to that of Bohr and Ulfbeck. In their words, "quantal physics thus emerges as but an implication of relativistic invariance, liberated from a substance to be quantized and a formalism to be interpreted" (1995, 1).

b.   it is not a theory of the behavior of matter-in-motion. Our ontology does not accept matter-in-motion as *fundamental* (though it is phenomenologically/pragmatically useful).

c.   so-called quantum entities and their characteristic properties such as entanglement are geometric features of the spacetime structure just as length contraction, on the *Minkowski-geometrical interpretation of special relativity*, is taken to be a feature of the geometry and not ultimately explained by the "inner constitution" of material bodies themselves[20].

(iii)   Some take the deeper physical insight of relativity to be the *true* metaphysical equivalence of all possible foliations of the spacetime manifold. We take this to mean that consistency – *metaphysical* consistency – with relativity at least demands that all foliations of spacetime be considered *equally real*. Our geometrical quantum mechanics embraces such a radical democracy of foliations. In this way, we are pursuing an analogy between NRQM and what is called the "geometrical rendition by Minkowski of special relativity" (Brown 2005, vii).

(iv)   Spatiotemporal relations are the means by which all physical phenomena (including both quantum and classical "entities") are modeled, allowing for a natural transition from quantum to classical mechanics (including the transition from quantum to classical probabilities) as simply the transition from rarefied to dense collections of spacetime relations.

(v)   Given (i) – (iv), we adopt an explanatory strategy that is faithful to our methodological and ontological commitments: we take the view that the determination of events, properties, experimental outcomes, etc., in spacetime is made with spacetime symmetries both *globally* and a*causally*. That is, we will invoke an acausal global determination relation that respects neither past nor future common cause principles. We will apply this methodology to a specific quantum mechanical set-up in section 7.

(vi)   As will be demonstrated in section 7, the reality of all events is necessary for explanation on our view, the blockworld assumption thus plays a *non-trivial explanatory role*.

*3.5 Motivating our geometrical interpretation of quantum theory.* In order to appreciate how we came to this view, we will outline our broad motivations for this brand of geometrical quantum mechanics. Our primary philosophical motivations, which have profound methodological implications for how one would model reality, are to eliminate various "dualisms" that currently plague theoretical physics. One main dualism is the following: "inner constitution of material bodies" vs. "their spatiotemporal background." For example, as long as one maintains this dualism, troubling questions such as the following will arise:

---

[20] A note on this explanatory strategy. It is rather controversial to claim that, on the Minkowski interpretation of STR, length contraction can be *explained*. This is because it is thought that a *pure* geometry of spacetime does not have the explanatory resources to say *why* it is that rods are the way they are; a pure geometry can *merely represent* the rod's *behavior* from different points of view in spacetime. However, we are here rejecting the fundamentality of constructive explanations in favor of principle geometric explanation. This is where our *ontology* of relations and the *global determination* of events with spacetime *symmetries* are important; see points (iv) and (v).

> if it is the structure of the background spacetime that accounts for the phenomenon [such as length contraction], by what mechanism is the rod or clock informed as to what this structure is? How does this material object get to know which type of spacetime – Galilean or Minkowskian, say – it is immersed in (Brown 2005, 8).

This may also be called the "matter-geometry" dualism.

There are certain constructive accounts of NRQM (e.g., collapse accounts such as the wavefunction view of GRW, or modal accounts such as Bohmian mechanics) where, if this dualism is true, you are led to a dilemma between the dynamics of NRQM state-evolution and kinematical coordinate transformations. So, here is the problem. One tries to interpret NRQM constructively, as a theory of the dynamics of matter-in-motion. And then, one tries to relate that theory to a *principle* account of spacetime structure where we take the kinematical transformations as simply perspectives on already-existing events and *independent of dynamical considerations*. But now, we are forced to either: (i) conclude that the *dynamical laws of motion* are in some sense wrong (i.e., that they are *not* invariant under a kinematical coordinate transformation) or (ii) that the *space* in which the matter-in-motion evolves has been entirely misconstrued (i.e., that we are not relating foliations of a spacetime with quantum objects *there*, but are relating the dynamical evolution of a quantum mechanical wavefunction in configuration space, from which we must extract an *image* of ordinary spacetime)[21].

For Brown, the solution to this conundrum is to *collapse* the fundamental distinction between kinematics and dynamics in favor of a dynamical account of spacetime structure from which one can reconstruct the essential features of the kinematical coordinate transformations on the basis of the ontology/dynamics supplied (via quantum theory, for example). Thus, with the appropriate underwriting story of spacetime structure in hand, one can derive the necessary coordinate transformations on the basis of how matter behaves. And with this, empirical adequacy is achieved.

As we will argue, our geometrical quantum mechanics with spacetime relations collapses the matter-geometry dualism and therefore avoids this dilemma without having to deny that kinematics and dynamics are conceptually distinct. We therefore embrace and defend a non-dynamical view of spacetime structure, *contra* Brown.

Given our geometric view of NRQM, we reject realism about the Hilbert space, for as David Albert says,

> the space in which any realistic interpretation of quantum mechanics is necessarily going to depict the history of the world as *playing itself out* … is *configuration*-space. And whatever impression we have to the contrary (whatever impression we have, say, of living in a three-dimensional space, or in a four-dimensional space) is somehow flatly illusory (1996, 277).

Given that spatiotemporal relations are fundamental on our view, we want no part of any interpretation that is embroiled with the problem of how to extract an image of a three-dimensional world from either the instantaneous state or the evolving state of a 3$N$- dimensional system. Moreover, we want to avoid any concerns about the *ontological status* of configuration space. This paper constitutes an extended defense of the claim that nothing about quantum mechanics *requires* denying the truth of 4D-ism, and it provides an interpretation of both STR and NRQM which is realist about 4-space and anti-realist about Hilbert space.

---

[21] For a discussion of this dilemma, *see* Lewis (2004), section 5.

In short, the geometrical perspective adopted here is inimical to: (a) theories which invoke a preferred frame for their dynamics (such as a neo-Lorentzian account of STR), (b) constructive accounts of either STR or NRQM, (c) any realistic interpretation of Hilbert space, and (d) accounts of NRQM for which the role of spacetime as a unifying descriptive framework, such as found in Minkowski's interpretation of STR, is either unclear or problematic (such as "many-worlds" interpretations of Everettian NRQM).

Many will assume that a geometric interpretation such as ours is impossible because quantum wavefunctions live in Hilbert space and contain much more information than can be represented in a classical space of three dimensions. The existence of entangled quantum systems provides one obvious example of the fact that more information is contained in the structure of quantum mechanics than can be represented completely in spacetime. As Peter Lewis says, "the inescapable conclusion for the wavefunction realist seems to be that the world has $3N$ dimensions; and the immediate problem this raises is explaining how this conclusion is consistent with our experience of a three-dimensional world" (2004, 717). On the contrary, the existence of the non-commutativity of quantum mechanics is deeply related to the structure of *spacetime* itself, without having to invoke the geometry of Hilbert space. Surprisingly, as will be demonstrated in the following section, it is a spacetime structure for which the relativity of simultaneity is upheld, and not challenged

## 4. The Relativity of Simultaneity and Non-relativistic Quantum Mechanics

Notice in the BW example from section 2 that if Joe 'jumps into' Sara's frame of reference at event 1 then moves spatially to Kim (Figure 6), he does not get to event 2 because he will be at the T = 0 version of Kim while the Kim of event 2 is at T = -0.0025s. If, rather, Joe moves spatially to Bob *then* 'jumps into' the girls' frame of reference, he is at the T = -0.0025s version of Kim. That is to say, Lorentz boosts (changes to moving frames of reference according to the Poincaré group of STR) do not commute with spatial translations since different results obtain when the order of these two operations is reversed. Specifically, this difference is a temporal displacement which is key to generating a BW.

This is distinct from Newtonian mechanics whereby time and simultaneity are absolute per Galilean invariance. If spacetime was Galilean invariant, the boys and girls in our example would all agree as to which events were simultaneous and could subscribe to presentism. In such a spacetime, it would not matter if you Galilean boosted then spatially translated, or spatially translated then Galilean boosted. *Prima facie*, one might suspect that *non-relativistic* quantum mechanics would be in accord with Galilean spacetime. And indeed, the linear dynamics – the Schrödinger equation – is Galilean invariant (Brown and Holland 1999). However, as we will show, while it is indeed true that the Schrödinger dynamics is Galilean invariant, the appropriate spacetime structure for which one can obtain the Heisenberg commutation relations is *not* a Galilean spacetime! Surprisingly, it is a spacetime structure "in between" Galilean spacetime and Minkowski spacetime, but one for which the relativity of simultaneity is upheld, unlike in Galilean spacetime.

Inevitably, the very means by which we can establish a determinate position in spacetime – or a determinate momentum (mass times velocity) – is going to have to speak to the quantum theory, a theory which places strictures on such questions. Now, a position can be given by an "axis of rotation" in a spacetime (just imagine a line around which some reference frame is spinning, or around which every other coordinate system is contracting if we are talking about Lorentz boosting from one frame to another). Such a thing can be picked out by "boost" operators, to use the language of the spacetime symmetry group. Given a Lorentz boost, one effectively picks out a position in spacetime (since the new coordinate systems given by the boost operator *all* share exactly their *origin* in common – thus uniquely picking out *one point in 2D spacetime and a line in 3D spacetime, etc.*). That is, the axis of rotation yields a spacetime trajectory which would yield a point in 'space' at any given time. Similarly, we might think about "momentum" as nothing but (speaking again in terms of spacetime groups) a scalar multiple of the generators of spatial translations. That is, spatially translating is simply "moving" from one position to another (albeit into a new frame); and this is something like a velocity (i.e., a time-derivative of position).

Now, if we *define* a commutator between position and momentum *in terms of* the generators of boosts and spatial translations respectively[22] – and note that they *do not commute* when simultaneity is nonabsolute (relative) – is it possible to show that one can arrive at the quantum-mechanical commutator *of position and momentum*, and have it *equal* to the quantum mechanically well-known quantity $-i\hbar$? This is equivalent to asking "what is the spacetime structure such that, if simultaneity is non-absolute, the Heisenberg commutator can be deduced?[23]"

Quite surprisingly, it turns out that *because* boosts do not commute with spatial translations *given that simultaneity is relative*, one can indeed deduce the quantum mechanical Heisenberg commutator (in the appropriate *"weakly"* relativistic spacetime regime). This shows that some interpretation exists for both non-relativistic quantum mechanics and any relativistic quantum mechanical theory, where there is a *single, unified spacetime arena* from which either theory can be obtained in the appropriate asymptotic limit. More specifically, what the formal results in the following sections will show is that classical mechanics "lives in" G4, surprisingly NRQM "lives in" a spacetime regime that is between G4 and $\mathcal{M}^4$ (we can call it K4 after Kaiser) and RQFT "lives in" $\mathcal{M}^4$. It will also become clear that NRQM is truly "baby" RQFT in that it also is about new trajectories—or particle creation to use dynamical lingo. All of this makes for a great deal more unity between spacetime structures and quantum structures than is generally appreciated.

---

[22] *See* equations (4.0.1) and (4.0.2) in section 4.1.

[23] And since quantum theory is already well-established empirically, we essentially know what needs to be derived, we just as-yet have not found the right spacetime structure. This is, admittedly, flipping the order of discovery somewhat, and asking an entirely new question regarding the "origin" of quantum theory (looking to spacetime structure, and not to the structure of matter per se, which is how the theory of the quantum was arrived at historically).

*4.1 NRQM: Spacetime structure for commutation relations.* Kaiser[24] has shown that the non-commutivity of Lorentz boosts with spatial translations is *responsible for* the non-commutivity of the quantum mechanical position operator with the quantum mechanical momentum operator. He writes[25],

> For had we begun with Newtonian spacetime, we would have the Galilean group instead of [the restricted Poincaré group]. Since Galilean boosts commute with spatial translations (time being absolute), the brackets between the corresponding generators vanish, hence no canonical commutation relations (CCR)! In the [c $\rightarrow \infty$ limit of the Poincaré algebra], *the CCR are a remnant of relativistic invariance where, due to the nonabsolute nature of simultaneity, spatial translations do not commute with pure Lorentz transformations.* [Italics in original].

Bohr & Ulfbeck[26] also realized that the "Galilean transformation in the weakly relativistic regime" is needed to construct a position operator for NRQM, and this transformation "includes the departure from simultaneity, which is part of relativistic invariance." Specifically, they note that the commutator between a "weakly relativistic" boost and a spatial translation results in "a time displacement," which is crucial to the relativity of simultaneity. Thus they write[27],

> "For ourselves, an important point that had for long been an obstacle, was the realization that the position of a particle, which is a basic element of nonrelativistic quantum mechanics, requires the link between space and time of relativistic invariance."

So, *the essence of non-relativistic quantum mechanics – its canonical commutation relations – is entailed by the relativity of simultaneity.*

　　　If the transformation equations entailed by some spacetime structure necessitate a temporal displacement when boosting between frames, then the relativity of simultaneity is true of that spacetime structure. Given this temporal displacement between boosted frames, and given that this implies the relativity of simultaneity, our arguments supplied above show that BW is true of this spacetime structure. Furthermore, since the relativity of simultaneity, via the kind of temporal displacement necessitated by boosting between frames in this spacetime regime, is essential to the Heisenberg or canonical commutation relations, we find a heretofore unappreciated deep unity between STR and *non-relativistic* quantum mechanics.

---

[24] Kaiser (1981 & 1990).
[25] Kaiser (1981), p. 706.
[26] Bohr & Ulfbeck (1995), section D of part IV, p. 28.
[27] *Ibid.*, p. 24.

To outline Kaiser's result, we take the limit c → ∞ in the Lie algebra of the Poincaré group for which the non-zero brackets are:

$$[J_m, J_n] = i J_k$$
$$[T_0, K_n] = i T_n$$
$$[K_m, K_n] = \frac{-i}{c^2} J_k$$
$$[J_m, K_n] = i K_k$$
$$[J_m, T_n] = i T_k$$
$$[T_m, K_n] = \frac{-i}{c^2} \delta_{mn} T_0$$

where expressions with subscripts m,n and k denote 1, 2 and 3 cyclic, $J_m$ are the generators of spatial rotations, $T_0$ is the generator of time translations, $T_m$ are the generators of spatial translations, $K_m$ are the boost generators, $i^2 = -1$, and c is the speed of light. We obtain

$$[J_m, J_n] = i J_k$$
$$[M, K_n] = 0$$
$$[K_m, K_n] = 0$$
$$[J_m, K_n] = i K_k$$
$$[J_m, T_n] = i T_k$$
$$[T_m, K_n] = \frac{-i}{\hbar} \delta_{mn} M$$

where M is obtained from the mass-squared operator in the c → ∞ limit since

$$c^{-2} \hbar T_0 = c^{-2} P_0$$

and

$$\frac{P_o}{c^2} = \left( M^2 + c^{-2} P^2 \right)^{1/2} = M + \frac{P^2}{2Mc^2} + O(c^{-4}).$$

Thus, $c^{-2} T_0 \to \frac{M}{\hbar}$ in the limit c → ∞. [M ≡ mI, where m is identified as "mass" by choice of 'scaling factor' ħ.] So, letting

$$P_m \equiv \hbar T_m \tag{4.1}$$

and

$$Q_n \equiv \frac{-\hbar}{m} K_n \tag{4.2}$$

we have:

$$\left[P_m, Q_n\right] = \frac{-\hbar^2}{m}\left[T_m, K_n\right] = \left(\frac{-\hbar^2}{m}\right)\left(\frac{i}{\hbar}\right)\delta_{mn} mI = -i\hbar\delta_{mn} I \qquad (4.3)$$

Bohr & Ulfbeck (1995) point out that in this "weakly relativistic regime" the coordinate transformations now look like:

$$X = x - vt$$
$$T = t - \frac{vx}{c^2} \qquad (4.4)$$

These transformations differ from Lorentz transformations because they lack the factor

$$\gamma = \left(1 - \frac{v^2}{c^2}\right)^{-1/2}$$

which is responsible for time dilation and length contraction. And, these transformations differ from Galilean transformations by the temporal displacement $vx/c^2$ which is responsible for the relativity of simultaneity, i.e., in a Galilean transformation time is absolute so T = t. Therefore, the spacetime structure of Kaiser *et. al.* lies between Galilean spacetime and Minkowski spacetime and we see that the Heisenberg commutation relations are not the result of Galilean invariance, where spatial translations commute with boosts, but rather they result from the relativity of simultaneity per Lorentz invariance.

*4.2 Heterodoxy: NRQM Does Not Live In Galilean Spacetime.* The received view has it that Schrödinger's equation is Galilean invariant, so it is generally understood that NRQM resides in Galilean spacetime and therefore respects absolute simultaneity[28]. However, as we have seen above, Kaiser (1981), Bohr & Ulfbeck (1995) and Anandan (2003) have shown independently that the Heisenberg commutation relations of NRQM follow from the relativity of simultaneity[29]. *Prima facie* these results seem incompatible with the received view, so to demonstrate that these results are indeed compatible, we now show that these results do not effect the Schrödinger dynamics[30].

　　Why is it that the *dynamics* of NRQM, given by the Schrödinger equation, are Galilean invariant? That is, why are the dynamics of NRQM unaffected by the relativity of simultaneity reflected in the geometry of Eq. 4.3.

　　To answer this question we operate on |ψ> first with the spatial translation operator then the boost operator and compare that outcome to the reverse order of operations. The spatial translation (by *a*) and boost (by v) operators in x are:

---

[28] *See* Brown and Holland (1999).
[29] Of course, all other commutation relations in NRQM follow from those of position and momentum – with the exception of spin. Since, operationally, spin measurements are simply binary outcomes in space related to, for example, the spatial orientation of a Stern-Gerlach apparatus, our model encompasses such properties as spin to the extent that we model all outcomes in space and time as *irreducible relations* between the spatiotemporal regions corresponding to source and detector.
[30] See also Lepore (1960) who also realizes that this time-shift between frames is without effect on the dynamics of Schrödinger evolution.

$$U_T = e^{-iaT_x} \quad \text{and} \quad U_K = e^{-ivK_x} \tag{4.5}$$

respectively. These yield:

$$U_K U_T \big| \psi \big\rangle = U_T U_K e^{iavmI/\hbar} \big| \psi \big\rangle \tag{4.6}$$

Thus, we see that the geometric structure of Eq. 4.3 introduces a mere phase to |ψ> and is therefore without consequence in the computation of expectation values. And in fact, this phase is consistent with that under which the Schrödinger equation is shown to be Galilean invariant[31].

Therefore, we realize that the spacetime structure for NRQM, while not $\mathcal{M}^4$ in that it lacks time dilation and length contraction, nonetheless contains a "footprint of relativity[32]" due to the relativity of simultaneity. Thus, there is an unexpected and unexplored connection between the relativity of simultaneity and the non-commutativity of NRQM. In light of this result, it should be clear that there is no metaphysical tension between STR and NRQM. This formal result gives us motivation for believing that NRQM is intimately connected to the geometry of (a suitable) spacetime[33].

*4.3 Philosophical significance.* One important point should be brought out, which reveals how we understand the relationship between spacetime structure (given by relativity) and the theory of quantum mechanics (in a non-Minkowskian, but non-Galilean, spacetime regime, i.e., K4). Most natural philosophers agree that STR just constrains the set of possible dynamical theories to those which satisfy the light and relativity postulates. It is often worried, as we have pointed out, that somehow quantum theory *violates* those constraints. The view we adopt here is importantly different, in that we distinguish between:

      (a) the question of how to relate the *structures of* quantum theory and relativity
      (b) the question of the compatibility of constructive interpretations of quantum theory and whether they violate relativistic constraints.

We interpret quantum theory as a theory of principle – detached from a *constructive* interpretation of it. We then point out, using a collection of formal results, that the spacetime structure for which one can obtain the Heisenberg commutation relations is one where the relativity of simultaneity is *upheld* – a fact often not appreciated in most interpretations of quantum theory. Furthermore, with an ontology of spacetime relations, we show how one can motivate and derive the Born rule, and to construct a quantum density operator from the spacetime symmetry group of any quantum experimental configuration, and how one can use this to deduce and then explain the phenomenon of quantum interference – all by appealing to nothing more than a spacetime structure for which one can obtain the Heisenberg commutator while obeying the relativity of simultaneity.

---

[31] *See* Eq. 6 in Brown and Holland (1999). A derivation of Eq. 4.3, assuming the acceptability of a phase difference such as that in Eq. 4.6, is in Ballentine (1990), p. 49 – 58.

[32] This phrase was used by Harvey Brown in a conversation with the authors while describing his work with Peter Holland (Brown and Holland, 1999).

[33] The Bohr *et. al.* result of section 5 below shows how to relate this spacetime geometry to non-relativistic quantum mechanics by showing how a quantum density operator can be constructed on the basis of the spacetime symmetry group of any quantum mechanical experiment with the appropriate symmetries.

We take the deepest significance of the Kaiser *et. al.* results to be that, given the asymptotic relationship between the spacetime structure of special relativity and the "weakly relativistic" spacetime structure of quantum theory, non-relativistic quantum mechanics is something like a relativity theory in an "embryonic" stage. It is "embryonic" in that it is yet without the Lorentz-contraction factor $\gamma$ that appears in the familiar Lorentz transformation equations of special relativity[34].

Having identified the appropriate spacetime structure for the Heisenberg commutation relations, and having discovered that this structure upholds the relativity of simultaneity, we have provided a principle explanation for the quantum. A natural question now arises: what would the appropriate description of NRQM and quantum mechanical phenomena such as interference be like in light of the asymptotic relationship between relativity and quantum theory? Our "geometric" interpretation of NRQM elaborated below is one answer to this question, an answer grounded in our fundamental ontology of spacetime relations.

## 5. Density Matrix Obtained via Symmetry Group

Having found which *spacetime* structure is appropriate for the Heisenberg commutation relations (whose empirical manifestation is quantum interference), we now seek to address the question of how to model – *in spacetime and not in Hilbert space* – any quantum system which manifests quantum interference. That is, we are asking:

> *how can we describe a quantum system with nothing more than the geometry of spacetime, where the relativity of simultaneity and the non-commutivity of position and momentum obtain?*

The following formal results provide us with an answer to this question.

*5.1 Density Matrix Obtained via Spacetime Symmetry Group.* We present a pedagogical version of the appendix to Bohr, Mottelson and Ulfbeck (2004a) wherein they show the density matrix can be derived using only the irreducible representations of the symmetry group elements, $g \in G$. We begin with two theorems from Georgi

> The matrix elements of the unitary, irreducible representations of G are a complete orthonormal set for the vector space of the regular representation, or alternatively, for functions of $g \in G$ (1999, 14)

which gives

---

[34] And given that it is the contraction/dilation phenomena, characteristic of relativity, that motivates the introduction of the "field" as a unifying structural device, non-relativistic quantum mechanics in light of this new spacetime structure is simply relativity minus the "field."

If a hermitian operator, H, commutes with all the elements, D(g), of a representation of the group G, then you can choose the eigenstates of H to transform according to irreducible representations of G. If an irreducible representation appears only once in the Hilbert space, every state in the irreducible representation is an eigenstate of H with the same eigenvalue (*ibid.*, p. 25).

What we mean by "the symmetry group" is precisely that group G with which some observable H commutes (although, these elements may be identified without actually constructing H). Thus, the mean value of our hermitian operator H can be calculated using the density matrix obtained wholly by D(g) and <D(g)> for all g ∈ G. Observables such as H are simply 'along for the ride' so to speak.

To show how, in general, one may obtain the density matrix using only the irreducible representations[35] D(g) and their averages <D(g)>, we start with eqn. 1.68 of Georgi (*ibid.,*18)

$$\sum_g \frac{n_a}{N} \left[ D_a(g^{-1}) \right]_{kj} \left[ D_b(g) \right]_{lm} = \delta_{ab} \delta_{jl} \delta_{km}$$

where $n_a$ is the dimensionality of the irrep, $D_a$, and N is the group order. If we consider but one particular irrep, D, this reduces to the orthogonality relation (eqn. 1) of Bohr *et al.*

$$\sum_g \frac{n}{N} \left[ D(g^{-1}) \right]_{kj} \left[ D(g) \right]_{lm} = \delta_{jl} \delta_{km} \qquad (5.1)$$

where n is the dimension of the irrep. Now multiply by $[D(g')]_{jk}$ and sum over k and j to obtain

$$\sum_j \sum_k \sum_g \frac{n}{N} \left[ D(g^{-1}) \right]_{kj} \left[ D(g) \right]_{lm} \left[ D(g') \right]_{jk} = \sum_j \sum_k \delta_{jl} \delta_{km} \left[ D(g') \right]_{jk} = \left[ D(g') \right]_{lm}$$

The first sum on the LHS gives:

$$\sum_j \left[ D(g^{-1}) \right]_{kj} \left[ D(g') \right]_{jk} = \left[ D(g^{-1})D(g') \right]_{kk}$$

The sum over k then gives the trace of $D(g^{-1})D(g')$, so we have:

$$\frac{n}{N} \sum_g \left[ D(g) \right]_{lm} Tr\left\{ D(g^{-1})D(g') \right\} = \left[ D(g') \right]_{lm}$$

Dropping the subscripts we have eqn. 2 of Bohr *et al*:

$$\frac{n}{N} \sum_g D(g) \; Tr\left\{ D(g^{-1})D(g') \right\} = D(g') \; . \qquad (5.2)$$

If, in a particular experiment, we measure directly the click distributions associated with the various eigenvalues of a symmetry D(g), we obtain its average outcome, <D(g)>, i.e., eqn. 3 of Bohr *et al*:

---

[35] Hereafter, "irreps."

$$\langle D(g)\rangle = \sum_i \lambda_i \, p(\lambda_i) \qquad (5.3)$$

where $\lambda_i$ are the eigenvalues of D(g) and p($\lambda_i$) are the distribution frequencies for the observations of the various eigenvalues/outcomes.

In terms of averages, Bohr *et al.* eqn. 2 becomes:

$$\frac{n}{N}\sum_g \langle D(g)\rangle Tr\{D(g^{-1})D(g')\} = \langle D(g')\rangle \qquad (5.4)$$

which they number eqn. 4. Since we want the density matrix to satisfy the standard relation (Bohr *et al.* eqn. 5):

$$Tr\{\rho D(g')\} = \langle D(g')\rangle \qquad (5.5)$$

it must be the case that (Bohr *et al.* eqn 6):

$$\rho \equiv \frac{n}{N}\sum_g D(g^{-1})\langle D(g)\rangle \qquad (5.6)$$

That this density operator is hermitian follows from the fact that the symmetry operators are unitary. That is, $D(g^{-1}) = D^{\dagger}(g)$ implies $\langle D(g^{-1})\rangle = \langle D(g)\rangle^*$, thus:

$$\rho^+ = \frac{n}{N}\sum_g D^+(g^{-1})\langle D(g)\rangle^* = \frac{n}{N}\sum_g D(g)\langle D(g^{-1})\rangle = \frac{n}{N}\sum_g D(g^{-1})\langle D(g)\rangle = \rho \,.$$

[The second-to-last equality holds because we are summing over all g and for each g there exists $g^{-1}$.] So, the density operator of eqn. 5.6 will be hermitian and, therefore, its eigenvalues (probabilities) are guaranteed to be real. This is not necessarily the case for D(g), since we know only that they are unitary. However, *we need only associate detector clicks with the eigenvalues of D(g) and in this perspective one does not attribute an eigenvalue of D(g) to a property of some 'click-causing particle'.* Therefore, whether or not the eigenvalues of any particular D(g) are real or imaginary is of no ontological or empirical concern.

*5.2 Philosophical significance.* With the above formal result in hand, we can now provide a clear answer to the question posed at the beginning of this section:

> *any system with the spacetime symmetry group characterized by D(g) will yield the quantum mechanical density matrix.*

The methodological significance of the Bohr *et. al.* formal result is that any NRQM system may be described with the appropriate *spacetime* symmetry group. But the philosophical significance of this proof is more interesting, and one rooted in our ontological spacetime relationalism.

Our view is a form of ontological structural realism which holds that the features of our world picked out by STR and NRQM are structures; moreover, we think that the structures picked out by our most successful theories to date – spacetime theories – are geometrical structures. And those structures, if taken seriously, are, we posit, structures of spacetime *relations*. Furthermore, we see the quantum theory as providing a *further structural constraint* on the introduction of spacetime events. Isolated to an idealized model of "sources," "detectors," "mirrors," etc. (see figure 11 for an idealized interferometer), our ontology is that each and every "click" or "measurement event" observed in the detector region *is itself evidence of a spacetime relation between the source-detector.* So, while the "click" itself maybe regarded as a transtemporal or classical object, it is not "caused by" a structural entity such as a particle that is independent from the physical spacetime geometry of this entire measurement process and experimental set-up, rather, the *click itself* is a manifestation of spatiotemporal relations between elements of the experimental set-up. It is in this way, via our radical ontology of spacetime relations[36], that the essential features of quantum systems with interference can be described with features of the spacetime geometry *without appealing to features of the usual Hilbert space of quantum mechanical states*[37]. Given that our geometric interpretation collapses the matter/geometry dualism in favor of geometric structures and given our explanatory geometric description of the quantum, then our ontological structuralism is a principle theory.

Secondly, as will be demonstrated below, the Bohr *et al.* proof will allow us to show that the posit of a blockworld – the reality of all spacetime events, and hence on our ontology, of all spacetime relations constituting those events – does real explanatory work as will be demonstrated in section 7. While one can imagine quite trivial explanations of EPR-Bell correlations invoking the blockworld[38], the Bohr *et. al.* result will allow us to provide a non-trivial, geometric explanation for such quantum correlations.

Thirdly, as demonstrated below, the Bohr *et. al.* result provides the foundation for our distinctly *geometrical* ontological structuralist[39] interpretation of NRQM. This ontology is an ontology of spatiotemporal relations which are the means by which all physical phenomena (including both quantum and classical "entities") are modeled. Our relationalism allows for a natural transition from quantum to classical mechanics (including the transition from quantum to classical probabilities) as simply the transition from rarefied to dense collections of spacetime relations[40].

## 6. The Geometric Interpretation of NRQM

In order to motivate our relational approach to physical reality, consider first a rival interpretation of NRQM which is antithetical to the view we are developing here, Bohmian mechanics. Bohmian mechanics provides us with a classical-like picture of reality[41]. It begins by modeling the behavior of a classical-like particle whose velocity is determined, via "Bohm's equation" (i.e., the "guiding field"), by a wavefunction; the wavefunction evolves according to Schrödinger's equation (Maudlin 1994, 118). Such particles always have well-defined locations in spacetime, and their total Hamiltonian is constructed from both a non-classical quantum potential and classical potential fields. In a basic twin-slit experiment, a simple picture of the mechanism behind the interference pattern is provided: a particle is directed deterministically by the guiding field to a particular location and registered as a "click" in a detector. Measurement on Bohm's theory is just like any other physical interaction. A constructive account of measurement, from particle to "click" registration, is provided by breaking down the whole process into particles and wavefunctions. A "click" is clearly the result of a causal process (however non-classical/non-local that process might be), and evidences a particle trajectory in spacetime.

---

[36] Which, if you want to speak constructively, "constitute" the spacetime geometry.

[37] A Hilbert space is not analogous to spacetime geometry, but rather to phase-space geometry. Anandan (1991) for example adopts the view that the geometry of Hilbert space is appropriate for a geometric interpretation of quantum theory.

[38] *E.g.* Barrett (2004) critiques one such trivial explanatory model, which he calls a "teleological spacetime map."

[39] *See* French & Ladyman 2003*a* for an account of ontological strucuturalism in the context of quantum theory.

[40] Though a full explication and defense of this view is unfortunately beyond the scope of this paper.

Given our principle, geometrical interpretation of NRQM, it should be clear that we do not take detector events to be indicators of the trajectories of classical-like particles and wavefunctions, as in Bohm's mechanics. From our rejection of Hilbert space realism, for example, the wavefunction in Hilbert space does not determine our experiences in spacetime. To motivate Bohm's equation, one must believe that the wavefunction determines the velocity of particles (*ibid.*), and hence what the world looks like. Bohm's equation is therefore unwarranted on our view.

More generally, our explanation for the detector events is not going to appeal to dynamical objects and their equations of motions, or the forces acting on them. Rather, *our project is to model denumerable and discrete sets of events with a space and time of four dimensions as the basic geometry of the world*. Clicks evidence irreducible spatiotemporal relations between the source and the detector (think of twin-slit experiments[42]). Given that we are forced to take collections of relations and not trajectories as fundamental, we must *construct* those trajectories out of such relations. Therefore, in order to more fully capture the manner by which trajectories are inferred and constructed (for example from the exchange of "bosons") we assume that the fundamental constituents for modeling trajectories in spacetime are *relations* per Anandan (2002), i.e., elements of S×S where S is the spacetime manifold.

*6.1 Trajectories in Spacetime from Relations.* In keeping with our principle, geometric approach to quantum mechanics, we will restrict ourselves to the modeling of what is observed in the measurement process: the spatiotemporal location of discrete events which we call detector events. Let $s_i \in S$. To be consistent with the assumption that spatiotemporal relations are fundamental, we are assuming that the worldlines of detector events begin in spacetime at the second event entry, $s_2$, of the relation $(s_1,s_2)$ in question (the first entry, $s_1$, is presumably an event in the spacetime region of the source; more on that below). Accordingly, physicists are charged to find rules that will allow them to predict the locations and shapes of the trajectories of quantum objects in a designated spacetime region, i.e., the distribution of detector events whence a trajectory for the quantum object is inferred in the spacetime region of the detector. Of course, these rules exist in quantum and classical physics so we need to map our geometric ontology of spacetime relations to the relevant rules of quantum and classical physics, i.e., to NRQM and non-relativistic classical mechanics (CM).

---

[41] *See* Holland (1993) p. 26 and 81ff.; Barrett (1999) sections 5.2 – 5.6; and Maudlin (1994) p. 116ff. for the senses in which Bohm is classical-like.
[42] *See* figures 8 – 10.

*6.2 Distribution of Detector Events.* We begin by showing how NRQM and CM provide the rules for predicting the spacetime locations of detector events. Since we are dealing with NRQM and CM[43], we consider a single-particle source emitting at a slow enough rate that there exists no more than one non-relativistic "quantum object" in the space occupied by the detector at any given time. For those trials with multiple detector events, we will find that the events reside on a trajectory satisfying the classical equations of motion per the relevant spacetime boundary conditions. This follows by assumptions implicit in the experimental arrangement. First, the assumption of a "single-particle" is defined by a single trajectory so if detector events fell along more than one trajectory in some trial, we would believe that our source had emitted a second particle near enough to the first in time that, contrary to our initial assumption, we had two particles in the detector region at the same time (or that scattering or particle decay had taken place, contrary to our criteria for membership in the experiment, i.e., an empirical examination of NRQM rather than QFT). Second, the trajectory realized by the detector events will reside in the relevant (classical) Hamilton-Jacobi family of possible trajectories because that is how one obtains the properties of a "quantum object," such as mass and charge, required to solve the Schrödinger equation. Further, the trajectory of the quantum object will be uniquely determined (among the family of possibilities) by the first of the detector events per the continuity equation, whence trajectories do not intersect. Now, CM provides the shapes of the trajectories in the family of possibilities, but it does not provide a rule for predicting which trajectory will be realized by our "quantum object[44];" that task falls to NRQM via the probability density, $\psi^*\psi$.

Again, we are trying to predict the location and shape of a trajectory (inferred from a collection of detector events) in a particular spacetime region, i.e., the "detector" region, D, a subset of S. We hold that $\psi^*\psi$ pertains only to the *first event* in an n-event trial (at least for those that satisfy the experimental assumptions) because if $\psi^*\psi$ were intended to hold for the first *and* subsequent events, then the fact that subsequent events fall along a trajectory, being highly improbable in general, would force us to dismiss $\psi^*\psi$ on empirical grounds. Therefore, the first detector event of a multiple-event trial determines which trajectory exists in D for any given trial, and the distribution of first events is given probabilistically by NRQM[45]. Subsequent detector events in each trial will fall on the trajectory, determined by the first event, whose shape is given by CM.

To summarize, we are denying the standard claim that, as Anandan puts it, "the particle at any given time is described by a wave-function … which is a *complex-valued function of space*" (1992, 307). On our view, particles, qua "objects," are described by

---

[43] Rather than QFT and STR.

[44] This is also true of classical objects, but it is typically of no concern as classical objects are *ipso facto* never "screened off" so the experimenter can exercise total control over the initial conditions.

[45] We are embracing a typical assumption of statistical physics, i.e., experiments are repeatable and the probability outcomes are realized in the frequencies of the repeated trials. Thus, we have many equivalent samples of D (which includes the relevant spatiotemporal boundary conditions), one trial of the experiment in each sample, and the union of all these samples, containing but the first event of each trial, then approaches $\psi^*\psi$ as the number of samples/trials increases.

*trajectories in spacetime* but according to our interpretation *trajectories must themselves be constructed from spatiotemporal relations*. On our view, quantum theory is a principle theory that provides a description of "first events" – i.e., it provides *rules* for determining the probability that a detection event will occur in some (bounded) region of spacetime[46]. Each "first event" picks out a trajectory from all that are possible in a family of trajectories, and subsequent events lie along that trajectory which is described classically. Quantum theory just gives the probabilistic rule for predicting these first events *in spacetime*[47]. It is the "first events" that constitute our geometric analogue of particle creation in RQFT. Since each detector event evidences a spacetime relation, it is therefore useful to show how one obtains the Born rule on this basis.

*6.3 The Born Rule.* We conclude the formal presentation of our geometric understanding of NRQM with a geometric motivation of the Born rule. This argument is largely appropriated from Anandan (2002*b*), although variations have been introduced to accommodate our view.

We begin by noting that the totality of all relations in region D (whether they are possible per NRQM or not) form a featureless set (think of a block of marble which is to be chiseled into a sculpture). Therefore, our first task is to articulate the reason for our restricted outcomes space, i.e., the subsets of S×S with $s_2$ in D whence we may infer trajectories for the trials of the experiment. Here we modify Anandan's "heuristic principle M" (2002*b*, 419) to read:

> A necessary and sufficient condition for a set of relations to be admissible as outcomes in a trial of our experiment is that the set should conform to the spacetime symmetries inherent in the experiment.

A detector event evidences but the second element $s_2$ of a relation $(s_1,s_2)$ and without information concerning the first entry $s_1$ we cannot hope obtain a full geometric characterization of the experiment. Therefore, assuming the missing information is summarized in the Hamiltonian describing the experiment, M restricts the possible sets of relations in any given experiment to those which conform to the spacetime symmetries of the Hamiltonian[48]. Of course, NRQM and CM employ the same Hamiltonian and we know how spacetime symmetries are used in CM to establish deterministically the shapes of trajectories via Noether's theorem, so we just need to understand how M is used to obtain $\psi^*\psi$, thereby specifying the rule by which one and only one trajectory is realized in each trial[49].

---

[46] Principle theories are usually taken to provide constraints on the behavior of phenomena (Janssen 2002, 426). For our purposes, we take NRQM *qua* principle theory to provide *constraints on the distribution of events in spacetime*, as we suggested in sections 4 & 5.

[47] It should be noted that since our spacetime structure respects the relativity of simultaneity and first trajectory events are fundamentally distinct from subsequent trajectory events, trajectories must be time-like to avoid the temporal ordering ambiguity of space-like trajectories. Of course, NRQM satisfies this constraint *ipso facto*.

[48] This generalizes to QFT by including interaction Hamiltonians which introduce gauge symmetries. See Anandan (2002*b*, p. 424). *See also* Zee (2003), pp 84 – 85.

[49] For situations involving entangled particles, one trajectory per particle is realized.

In this context, we want to compute the probability density of finding the first event of a given trial in the neighborhood of $s_2$ in D. We require but one more conceptual-interpretative adjustment to Anandan's argument before we can appropriate its details for the origin of Born's rule in our view. In discussing the relation between $s_1$ in the source and $s_2$ in the detector, Anandan argues that all possible paths between $s_1$ and $s_2$ are equally probable per his assumption that "there are no causal dynamical laws." Specifically, if there existed a weighting of the various paths, this weighting would constitute a "causal dynamical law," albeit probabilistic, in violation of his assumption (Anandan 2002$b$, 425). Since we are working in the realm of relations rather than paths, we need to articulate the sense in which a path is a collection of relations. Of course, this decomposition is straightforward if we assume simply that relations are transitive, i.e., $(s_n,s_k) + (s_k,s_m) = (s_n,s_m)$. Thus, we assume[50] that all possible combinations of relations equivalent to $s_1$ in the source and $s_2$ in the detector are to be considered equally in computing the probability of a detector event in the neighborhood of $s_2$. Anandan's argument follows precisely from here.

If we start with the naïve assumption that the method by which all possible combinations of relations from $s_1$ to $s_2$ (or equivalently, "all possible paths") contribute to the probability outcome is via addition, we find the contribution from each path must be zero because there are an infinite number of such paths. To counter this result, without introducing an *ad hoc* weighting of paths, we need to have cancellation in the addition process. We therefore introduce a probability amplitude for each relation, such that the probability amplitude for a path should be constructed multiplicatively from the probability amplitudes of its relations per the transitivity of relations, i.e., the probability amplitude of $(s_n,s_k)$ times that of $(s_k,s_m)$ equals the probability amplitude of $(s_n,s_m)$. Then, the final probability for $(s_1,s_2)$ is found, by a means to be determined, after first adding the probability amplitudes of the equivalent paths. In this fashion, we might expect some cancellation in the addition process. To obtain a non-negative probability from an amplitude we need a norm "| |" over the amplitudes. That the probability amplitude of $(s_n,s_k)$, denoted by $\psi$, multiplied by that of $(s_k,s_m)$, denoted by $\varphi$, equals the probability amplitude of $(s_n,s_m)$ suggests $|\psi \varphi| = |\psi| |\varphi|$. Theorems by Hurwitz (1898) and Albert (1947, 495) state that these probability amplitudes should be reals, complex numbers, quaternions, or octonions (Adler 1995, 6 – 7; 109 – 111).

Octonions are not candidates for probability amplitudes since they are non-associative under multiplication when addition is also used, i.e., in general we do not have $|\psi_1(\psi_2 \, \psi_3) + \varphi| = |(\psi_1\psi_2) \, \psi_3 + \varphi|$. Reals are excluded because the only way to get cancellation between them is to use negative numbers, but the norms of negative numbers equal their positive counterparts so when working with an infinite number of paths we would still find the probability amplitude of each path is zero. Adler (*ibid.*) showed that it is not possible to construct a path integral using quaternions, so that leaves us with complex numbers for our probability amplitudes.

Now we find the probability for $(s_n,s_m)$ from the probability amplitudes for $(s_n,s_k)$, denoted by $\psi_1$, and $(s_k,s_m)$, denoted by $\psi_2$, in order to obtain the Born rule. If the phase of a relation is completely uncertain, then we expect the average of the probability for ($\psi_1 + \psi_2$) over all possible relative phases, ($\theta_1 - \theta_2$), will equal the sum of the individual probabilities for each of $\psi_1$ and $\psi_2$, i.e.,

$$\frac{1}{2\pi} \int_0^{2\pi} d\theta_1 P(\psi_1 + \psi_2) = P(\psi_1) + P(\psi_2) \tag{6.1}$$

where all possible relative phases are realized by having $\theta_1$ assume all values between zero and $2\pi$, $P(\psi_i)$ is the probability for the probability amplitude $\psi_i$ and $\theta_i$ is the phase of $\psi_i$. Since we have integrated over $\theta_1$, $P(\psi_1)$ on the right hand side of Eq. 6.1 is not a function of its phase, which means that in general $P(\psi)$ is a function of $|\psi|$ only. Since $P(\psi)$ is non-negative, it is reasonable to assume $P(\psi) = |\psi|^n$, where n is a non-negative integer. Now, that each path is equally likely means $P(\psi_1) = P(\psi_2)$, and therefore $|\psi_1| = |\psi_2|$. Let $|\psi_i| = b$ so that $|\psi_1 + \psi_2| = |[2b^2(1 + \cos(\theta_1 - \theta_2))]^{1/2}| = 2b|\cos(\theta/2)|$, where $\theta = \theta_1 - \theta_2$ and "| |" around terms containing the cosine functions means "absolute value."

---

[50] Of course, this statement is the geometric counterpart to the Feynman path integral formulation of NRQM.

Eq. 6.1 now reads:

$$\frac{2^n b^n}{2\pi} \int_0^{2\pi} d\theta \left| \cos^n(\theta/2) \right| = 2b^n$$

Letting $\alpha = \theta/2$ we have:

$$\frac{2^n b^n}{2\pi} 2\int_0^{\pi} d\alpha \left| \cos^n \alpha \right| = 2b^n$$

or

$$\frac{2^n}{2\pi} \int_0^{\pi} d\alpha \left| \cos^n \alpha \right| = 1 \tag{6.2}$$

A useful identity is:

$$\int_0^{\pi} d\alpha \left| \cos^{2m+1} \alpha \right| = 2 \int_0^{\pi/2} d\alpha \cos^{2m+1} \alpha \tag{6.3}$$

for any non-negative integer m. Evaluating the integral in Eq. 6.3 gives:

$$2 \int_0^{\pi/2} d\alpha \cos^{2m+1} \alpha = \frac{2^{2m+1}(m!)^2}{(2m+1)!} \tag{6.4}$$

Eqs. 6.2, 6.3 and 6.4, with n = 2m +1, give:

$$\frac{2^n}{2\pi}\frac{2^n}{n!}\left(\left(\frac{(n-1)}{2}\right)!\right)^2 = 1 \tag{6.5}$$

Since we need $\pi$ to cancel on the left hand side, (n-1) must be odd thus n must be even. Therefore, let n = 2m and Eq. 6.5 becomes:

$$\frac{2^{2m}}{2\pi}\frac{2^{2m}}{(2m)!}\left(\left(m-\frac{1}{2}\right)!\right)^2 = 1 \tag{6.6}$$

so we can use

$$\left(m-\frac{1}{2}\right)! = \frac{\sqrt{\pi}}{2^m}\frac{(2m)!}{2^m\, m!}$$

to cancel $\pi$ in Eq. 6.6, which then becomes

$$\frac{(2m)!}{2(m!)^2} = 1 \tag{6.7}$$

Eq. 6.7 holds for m = 1, but not for m > 1. Therefore, the only way Eq. 6.1 can hold with P($\psi$) = |$\psi$|$^n$ is to have n = 2, which is the Born rule.

### 6.4 Conclusion: Interpretive consequences of our geometrical NRQM.

*The Measurement Problem.* According to the account developed here, we offer a deflation of the measurement problem with a novel form of the "statistical interpretation." The fundamental difference between our version of this view and the usual understanding of it is the following: whereas on the usual view the state description refers to an "ensemble" which is an ideal collection of similarly prepared quantum particles, "ensemble" according to our view is just an ideal collection of spacetime regions $D_i$ "prepared" with the same spatiotemporal boundary conditions per the experimental configuration itself. The union of the first events in each $D_i$, as i → ∞, produces the characteristic Born distribution[51]. Accordingly, probability on our geometrical NRQM is interpreted per relative frequencies. It should be clear, also, that probabilities are understood as the likelihood that a particular relation between source-detector in spacetime is realized, from among a set of all equally likely relations between source-detector.

On our view, the wavefunction description of a quantum system can be interpreted statistically because we now understand that, as far as measurement outcomes are concerned, the Born distribution has a basis in the spacetime symmetries of experimental configurations. Each "click," which some would say corresponds to the impingement of a particle onto a measurement device and whose probability is computed from the wavefunction, corresponds to a spacetime relation in the context of the experimental configuration. The measurement problem *exploits* the possibility of extending the wavefunction description from the quantum system to the whole measurement apparatus, whereas the spacetime description according to our geometrical quantum mechanics *already includes* the apparatus via the spacetime symmetries instantiated by the *entire* experimental configuration. The measurement problem is therefore a non-starter on our view.

---

[51] There would be N first events in trials with N entangled particles, since each "particle" would correspond to a family of possible trajectories.

More importantly, following the Bohr *et. al.* results invoked throughout this paper, the spacetime symmetry group of an experimental configuration *entails* its density matrix. According to our view, the reason for the confusion over the ontological status of the wavefunction is illustrated nicely by the twin-slit experiment which "has in it the heart of quantum mechanics. In reality, it contains the *only* mystery" (Feynman *et al*, 1965, italics theirs). Starting, as in the Mach-Zehnder interferometer[52] (MZI) with a source and detector, we have a relationship that is described via spatial translation. If we illustrate this relationship via the *orbit of the translation operator* (figure 7), it is easy to see why one might infer the existence of transtemporal objects "emitted by the source and impinging on the detector." When one adds the double slit, the relationship established by the experimental configuration (source, slits, detector) involves a pair of translations between the source and each slit, and between each slit and the ultimate location of an event at the detector (figure 8). Therefore, the distribution of clicks at the detector is given by

$$\psi(\theta) = A\left(e^{ikx_1(\theta)} + e^{ikx_2(\theta)}\right)$$

which is, while itself not a translation, just the sum of spatial translations (figure 9).

It is easy to see why this event distribution is commonly attributed to "wave interference," especially with the addition of explicit time dependence[53] but the wavefunction has no fundamental, ontological status in spacetime. If many events are accumulated, the pattern will seem to add credence to an ontological interpretation of "wave interference." But

> the pattern is built one event (click) at a time and the explanation of each click is simply given by the appropriate composition of translations (figure 9).

Accordingly, there is no "wave" or "particle" emitted by the source, moving through the slits and impinging on the detector. *The key to deflating the mystery of wave-particle duality is that the orbits of the relevant spacetime symmetries are not worldlines.* Rather, according to our geometrical quantum mechanics, the orbits correspond to spacetime relations – relations between the source, detectors, beam splitter, etc. As shown in section 7, this is strikingly illustrated in the MZI where one sees the directions of orbits reversed under reflection so we have an orbit from source to BS1, from BS1 to the upper mirror, from the lower mirror to BS1, etc. (figure 11).

---

[52] *See* section 7 below, and figure 11.
[53] E.g. Shankar 1994.

The standard twin-slit configuration employs a source, screen with two slits and a detector surface (figures 8-10). Per the fundamentality of spacetime relations the screen reduces the collection of 'Huygens sources' at the screen's location to just two – one for each slit (figure 8). The relational result between these two 'Huygens sources' and the detector produces an interference pattern. When the experiment is conducted with electrons for example (Park, 1992, p 30), the interference pattern per the fundamentality of relations is realized. Using this result, we can illustrate a quantum-to-classical transition in twin-slit experiments by contrasting single-event trials with n-event trials.

Suppose we convert regions A and B (figure 10) to detector regions (*a la* the cloud chamber), referring to the original detector as the "final detector surface" to avoid confusion. For those trials in which the first event lies in region A, a classical trajectory is selected in region A so subsequent events do not contribute to an interference pattern at the final detector surface. That is, we establish a classical pattern in region A so we never have a quantum pattern in region B and these trials cannot provide a transition scenario. Therefore, we explore those trials in which the first event lies in region B.

There are two families of trajectories in region B, i.e., a family based at each slit. If the first event lies close to slit 1 (or 2), it is probable that the trajectory will be associated with family 1 (or 2). Therefore, the collection of trial-terminating events at the final detector surface in these trials will be in accord with trajectories emanating from slit 1 or 2 and terminating at the final surface without interference (classical case). If on the other hand the first event lies close to the final detector surface, the final event will also be close to the final surface. Since the first event must correspond to $\psi^*\psi$, the collection of trial-terminating events at the final detector surface in these trials will (artificially) evidence interference (quantum case). Therefore, a quantum-to-classical transition can be illustrated experimentally via the partition of all trials per the initial event position in region B – when the initial event is close to the slits, the distribution of events at the final detector surface is classical and when the initial event is close to the final detector surface, the distribution of events at the final detector surface is quantum.

Having argued, by way of appropriating Anandan's (1991; 2002) program, that complex probability amplitudes should be associated with all sets of spacetime relations equivalent to that relation for which we are trying to compute the probability of occurrence in the detector, we now understand the non-dynamical role of the Schrödinger equation and the wavefunction, which further distances our view from Anandan's. Schrödinger's equation is not describing the dynamic evolution of an entity in space, nor does it describe the "state of" a dynamic entity moving through space. Rather, Schrödinger's equation simply provides a calculation of the probability amplitudes for what *does* exist at the most fundamental level, spacetime relations.

*Entanglement & Non-locality.* On our geometric view of NRQM we explain entanglement as a feature of the spacetime geometry[54] as follows. Each initial detection event, which evidences a spacetime relation, selects a trajectory from a family of possible trajectories (one family per entangled 'particle'). In the language of detection events *qua* relations, it follows that correlations are correlations between the members of the *families* of trajectories and these correlations are the result of the relevant spacetime symmetries for the experimental configuration. And, since an experiment's spacetime symmetries are manifested in the Hamilton-Jacobi families of trajectories throughout the relevant spacetime region D, there is no reason to expect entanglement to diminish with distance from the source. Thus, the entanglement of families of trajectories is spatiotemporally global, i.e., non-local. That is, there is no reason to expect entanglement geometrically construed to respect any kind of common cause principle. Obviously, on our geometric interpretation there is no non-locality in the odious sense we find in Bohm for example, that is, there are no instantaneous causal connections (construed dynamically or in terms of production—bringing new states of affairs into being) between space-like separated events.

---

[54] Established in section 3 as one which is "weakly" relativistic in that it lacks the Lorentz contraction factor.

Quantum non-locality and entanglement are demystified in a straightforward fashion since spatiotemporal relations are fundamental in a blockworld. Correlations between space-like separated events that violate Bell's inequalities are of no concern as long as spatiotemporal relations in the experimental apparatus warrant the correlations. There is no need to satisfy either past or future versions of the common cause principle, since non-local correlations are not about "particles" impinging on measuring devices or what have you. Rather, the non-local correlations derive from the spatiotemporal relations in the construct of the experiment. There are no influences, causal mechanisms, etc., because non-locality is a relational property that is precisely described by the spacetime symmetries of any given experimental arrangement.

We understand quantum facts to be facts about the spatiotemporal relations of a given physical system, not facts about the behavior of particles, or the interactions of measurement devices with wavefunctions, or the like. Entanglement and non-locality are built into the structure of spacetime itself via relations. Correlations between space-like separated events that violate Bell's inequalities are of no concern as long as spacetime symmetries instantiated by the experimental apparatus warrant the correlated spacetime relations. Since the non-local correlations derive from the spatiotemporal relations per the spacetime symmetries of the experiment, satisfaction of any common-cause principle is superfluous. To sloganize: ours is a *purely* geometric/spacetime interpretation of non-relativistic quantum mechanics.

That the density matrix may be obtained from the spacetime symmetries of the Hamiltonian is consistent with the notion that $\psi^*\psi$ provides the distribution for detector events in single-event trials for each family of trajectories obtained via the Hamilton-Jacobi formalism. Our view exploits this correspondence to infer the existence of a single spacetime relation between source and detector for each detector event.

Subsequent detector events in close spatiotemporal proximity to the first tend to fall along a trajectory of the family consistent with the first event thereby allowing for the inference of a "particle." In this sense, what constitutes a "rarefied" distribution of spacetime relations is but one relation per "particle," i.e., family of trajectories, since subsequent events tend to trace out classical trajectories (scattering and particle decay events aside). It is a collection of these single-event trials that will evidence quantum interference in the twin-slit experiment.

**7. Application to "Quantum Liar Experiment"**

We will now apply the Bohr *et. al.* method to a particular experimental set-up. In two recent articles, Elitzur and Dolev try to establish something like the negation of the blockworld view, by arguing for an intrinsic direction of time given by the dynamical laws of quantum theory (2005 *a* & *b*). They put forward the strong claim that certain experimental set-ups such as the quantum liar experiment[55] "entail inconsistent histories" that "undermine the notion of a fixed spacetime within which all events maintain simple causal relations. Rather, it seems that quantum measurement can sometimes 'rewrite' a process's history" (2005*b*, 593). In response, they propose a "spacetime dynamics theory" (2005*a*, 345). Certainly, if something like this is true, then blockworld is jeopardized. By applying the geometrical interpretation of quantum mechanics to the "quantum liar" case, we will not only show that the blockworld assumption is consistent with such experiments, but that blockworld *a la* our geometric interpretation provides a non-trivial and unexpected explanation of such experiments.

The history of QM is littered with comparatively radical or reactionary attempts to explain features such as EPR-Bell correlations. For example, some accounts of NRQM give up the (past) common cause principle and invoke some kind of backwards-causal theory to explain quantum phenomena (Price, 1996). Others argue that EPR-Bell correlations require no (causal) explanation whatsoever (Fine, 1989). We provide another interpretation, one which is neither causal in character, nor merely skeptical about the possibility of causal explanations of EPR-like phenomena – but which is genuinely *acausal* and deeply revelatory about the origin of both the theory of quantum mechanics and its seemingly mysterious class of phenomena. Our geometric interpretation and its explanatory methodology reject both past and future versions of the common cause principle.

Our account provides a clear description, in terms of fundamental spacetime relations, of quantum phenomena *that does not suggest the need for a "deeper"* causal or dynamical *explanation*. If explanation is simply determination, then our view explains the structure of quantum correlations by invoking what can be called *acausal global determination relations*. These global determination relations are given by the spacetime symmetries which underlie a particular experimental set-up. Not objects and dynamical laws, but rather acausal spacetime relations per the relevant spacetime symmetries do the fundamental explanatory work according to our version of geometrical quantum mechanics.

*7.1 Mach-Zehnder Interferometer & Interaction Free Measurements.* Since QLE employs interaction-free measurement[56] (Elitzur & Vaidman, 1993), we begin with an explication of IFM. Our treatment of IFM involves a simple MZI (figure 11). All photons in this configuration are detected at D1 since the path to D2 is ruled out by destructive interference. This obtains even if the MZI never contains more than one photon in which case each photon "interferes with itself." If we add a detector D3 along either path (figures 12a and 12b), we can obtain clicks in D2 since the destructive interference between BS2 and D2 has been destroyed by D3. If we introduce detectors along the upper and lower paths between the mirrors and BS2, obviously we do not obtain any detection events at D1 or D2.

---

[55] Hereafter, simply "QLE."
[56] "IFM."

To use this MZI for IFM, we place an atom with spin X+, say, into one of two boxes according to a Z spin measurement, i.e., finding the atom in the Z+ (or Z-) box means a Z measurement has produced a Z+ (or Z-) result. The boxes are opaque for the atom but transparent for photons in our MZI. Now we place the two boxes in our MZI so that the Z+ box resides in the lower arm of the MZI (figure 13). If we obtain a click at D2, we know that the lower arm of the MZI was blocked as in figure 12a, so the atom resides in the Z+ box. However, the photon must have taken the upper path in order to reach D2, so we have measured the Z component of the atom's spin without an interaction. Accordingly, the atom is in the Z+ spin state and subsequent measurements of X spin will yield X+ with a probability of one-half (whereas, we started with a probability of X+ being unity).

*7.2 Quantum Liar Experiment.* The QLE leads to the quantum liar paradox of Elitzur and Dolev (2005a, pp 325-50) because it presumably instantiates a situation isomorphic to a liar paradox such as the statement: "this sentence has never been written." As Elitzur and Dolev put it, the situation is one in which we have two distinct non-interacting atoms in different wings of the experiment that could only be entangled via the mutual interaction of a single photon. However one atom is found to have blocked the photon's path and thus it could not interact with the other atom via the photon and the other atom should therefore not be entangled with the atom that blocked the photon's path. But, by violating Bell's inequality, its "having blocked the photon" was affected by the measurement of the other atom, hence the paradox. Our explication of the paradox differs slightly in that we describe outcomes via spin measurements explicitly.

We start by exploiting IFM to entangle two atoms in an EPR state, even though the two atoms never interact with each other or the photon responsible for their entanglement (Hardy, 1992)[57]. We simply add another atom prepared as the first in boxes Z2+/Z2- and position these boxes so that the Z2- box resides in the upper arm of the MZI (figure 14). Of course if the atoms are in the Z1+/Z2- states, we have blocked both arms and obtain no clicks in D1 or D2. If the atoms are in Z1-/Z2+ states, we have blocked neither arm and we have an analog to figure 11 with all clicks in D1. We are not interested in these situations, but rather the situations of Z1+ *or* Z2- as evidenced by a D2 click. Thus, a D2 click entangles the atoms in the EPR state:

$$\frac{1}{\sqrt{2}}\left(\left|Z+\right\rangle_1 \left|Z+\right\rangle_2 + \left|Z-\right\rangle_1 \left|Z-\right\rangle_2\right)$$

and subsequent spin measurements with orientation of the Stern-Gerlach magnets in $\Re^2$ as shown in figure 15 will produce correlated results which violate Bell's inequality precisely as illustrated by Mermin's apparatus (Mermin, 1981). This EPR state can also be obtained using *distinct* sources as shown by Elitzur *et. al.* (2002; *see* figure 16), so a single source is not necessary to entangle the atoms. In either case, subsequent spin measurements on the entangled atoms will produce violations of Bell's inequality.

---

[57] The non-interaction of the photons and atoms is even more strongly suggested in an analogous experiment, where a super-sensitive bomb is placed in on of the arms of the MZI. *See* Aharonov & Rohrlich (2005), esp. section 6.4 and chapter 17.

Suppose we subject the atoms to spin measurements after all D2 clicks and check for correlations thereafter. A D2 click means that one (and only one) of the boxes in an arm of the MZI is acting as a "silent" detector, which establishes a "fact of the matter" as to its Z spin and, therefore, the other atom's Z spin. In all trials for which we chose to measure the Z spin of both atoms this fact is confirmed. But, when we amass the results from all trials (to include those in which we measured Γ and/or Δ spins) and check for correlations we find that Bell's inequality is violated, which indicates the Z component of spin *cannot* be inferred as "a matter of unknown fact" in trials prior to Γ and/or Δ measurements. This is not consistent with the apparent "matter of fact" that a "silent" detector must have existed in one of the MZI arms in order to obtain a D2 click, which entangled the atoms in the first place. To put the point more acutely, Elitzur and Dolev (2005a, 344) conclude their exposition of the paradox with the observation that

> *The very fact that one atom is positioned in a place that seems to preclude its interaction with the other atom leads to its being affected by that other atom.* This is logically equivalent to the statement: "This sentence has never been written.[58]"

In other words, *there must be a fact of the matter concerning the Z spins in order to produce a state in which certain measurements imply there is no fact of the matter for the Z spin.*

*7.3 Geometrical account of QLE.* By limiting any account of QLE to a story about the interactions of objects or entities in spacetime (such as the intersection of point-particle-worldlines, or an everywhere-continuous process *connecting* two or more worldlines), it is on the face of it difficult to account for "interaction-free" measurements (since, naively, a necessary condition for an "interaction" is the "intersection of two or more worldlines). Since the IFM in this experiment "generated" the entanglement, we can invoke the *entire* spacetime configuration of the experiment so as to predict, and explain, the EPR-Bell correlations in QLE. Indeed, it has been the purport of this paper that the spacetime symmetries of the quantum experiment can be used to construct its quantum density operator, that such a spacetime is one for which simultaneity is relative, that events in the detector regions evidence spatiotemporal relations, and that the Born rule can be *derived* on the basis of the geometry of spacetime relations.

Accordingly, spatiotemporal relations provide the ontological basis for our principle geometric interpretation of quantum theory, and on that basis, explanation (*qua* determination) of quantum phenomena can be offered. According to our ontology of relations, the distribution of clicks at the detectors reflects the spatiotemporal relationships between the source, beam splitters, mirrors, and detectors as described by the spacetime symmetry group – spatial translations and reflections in this case. The

---

[58] This quote has been slightly modified per correspondence with the authors to correct a publisher's typo. In the original document they go on to point out that "[we] are unaware of any other quantum mechanical experiment that demonstrates such inconsistency" (*ibid.*).

relevant 2D irreps for 1-dimensional translations and reflections are (Bohr & Ulfbeck, 1995):

$$T(a) = \begin{pmatrix} e^{-ika} & 0 \\ 0 & e^{ika} \end{pmatrix} \quad \text{and} \quad S(a) = \begin{pmatrix} 0 & e^{-2ika} \\ e^{2ika} & 0 \end{pmatrix}$$

respectively, in the eigenbasis of T. *These are the fundamental elements of our geometric description of the MZI.* Since, with this ontology of spatiotemporal relations, the matter-geometry dualism (as explained in section 2) has been collapsed, both "object" and "influence" reduce to *spacetime relations*. The entanglement found in this experimental arrangement reduces to the spatiotemporal relationship between two families of trajectories, one family for each 'atom' in subsequent spin measurements. We can then obtain the *density matrix* for such a system via its spacetime symmetry group (Bohr & Ulfbeck, 1995). Recall that the density matrix characterizes the "entanglement" now understand as entanglement between families of trajectories.

Consider now figure 11, with the present geometrical interpretation of quantum mechanics in mind. We must now re-characterize that experimental set-up in our new geometrical language, using the formalism of Bohr *et. al.* Let a detection at D1 correspond to the eigenvector |1> (associated with eigenvalue $e^{-ika}$) and a detection at D2 correspond to the eigenvector |2> (associated with eigenvalue $e^{ika}$). The source-detector combo alone is simply described by the click distribution |1>. The effect of introducing BS1 is to change the click distribution per the unitary operator (Bohr & Ulfbeck 1995)

$$Q(a_o) \equiv \frac{1}{\sqrt{2}} \left( I - iS(a_o) \right)$$

where $a_o \equiv \pi/(4k)$. Specifically,

$$Q(a_o) = \frac{1}{\sqrt{2}} \left[ \begin{pmatrix} 1 & 0 \\ 0 & 1 \end{pmatrix} - i \begin{pmatrix} 0 & -i \\ i & 0 \end{pmatrix} \right] = \frac{1}{\sqrt{2}} \begin{pmatrix} 1 & -1 \\ 1 & 1 \end{pmatrix}$$

and

$$|\psi\rangle = Q(a_o)|1\rangle = \frac{1}{\sqrt{2}} \begin{pmatrix} 1 & -1 \\ 1 & 1 \end{pmatrix} \begin{pmatrix} 1 \\ 0 \end{pmatrix} = \frac{1}{\sqrt{2}} \begin{pmatrix} 1 \\ 1 \end{pmatrix}$$

This is an eigenstate of the reflection operator, so introducing the mirrors does not change the click distribution. Introduction of the second beam splitter, BS2, changes the distribution of clicks at D1 and D2 per

$$|\psi_{final}\rangle = Q^+(a_o)|\psi\rangle = \frac{1}{\sqrt{2}} \begin{pmatrix} 1 & 1 \\ -1 & 1 \end{pmatrix} \begin{pmatrix} \dfrac{1}{\sqrt{2}} \\ \dfrac{1}{\sqrt{2}} \end{pmatrix} = \begin{pmatrix} 1 \\ 0 \end{pmatrix}$$

Note there is no mention of photon interference here. We are simply describing the distribution of events (clicks) in spacetime (spatial projection, rest frame of MZI) using the fundamental ingredients in this type of explanation, i.e., spacetime symmetries (spatial translations and reflections in the MZI, rotations in the case of spin measurements). What it means to "explain" a phenomenon in this context is to provide the distribution of spacetime events per the spacetime symmetries relevant to the experimental configuration.

To complete our geometrical explanation of QLE we simply introduce another detector (D3 as in figure 12a, say), which changes the MZI description *supra* prior to BS2 in that the distribution of clicks for the configuration is given by

$$\left| \psi_{final} \right\rangle = \begin{pmatrix} Q^+(a_o) & & 0 \\ & \ddots & 0 \\ 0 & 0 & 1 \end{pmatrix} \begin{pmatrix} 1/\sqrt{2} \\ 0 \\ 1/\sqrt{2} \end{pmatrix} = \begin{pmatrix} 1/\sqrt{2} & 1/\sqrt{2} & 0 \\ -1/\sqrt{2} & 1/\sqrt{2} & 0 \\ 0 & 0 & 1 \end{pmatrix} \begin{pmatrix} 1/\sqrt{2} \\ 0 \\ 1/\sqrt{2} \end{pmatrix} = \begin{pmatrix} 1/2 \\ -1/2 \\ 1/\sqrt{2} \end{pmatrix}$$

Again, we need nothing more than $Q^+$, which is a function of S(a), to construct this distribution. And for the distribution of clicks for the configuration in figure 12b

$$\left| \psi_{final} \right\rangle = \begin{pmatrix} 1/\sqrt{2} & 1/\sqrt{2} & 0 \\ -1/\sqrt{2} & 1/\sqrt{2} & 0 \\ 0 & 0 & 1 \end{pmatrix} \begin{pmatrix} 0 \\ 1/\sqrt{2} \\ 1/\sqrt{2} \end{pmatrix} = \begin{pmatrix} 1/2 \\ 1/2 \\ 1/\sqrt{2} \end{pmatrix}$$

Of course, spin measurements using the MZI boxes ("spin measurements on the atoms") are viewed as binary outcomes in space (spin ½) with respect to the orientation of the magnetic poles in a Stern-Gerlach device (SG). This is "how the atom was placed in the boxes according to spin." Successive spin measurements are described via rotation, i.e.,

$$\left| \psi_2 \right\rangle = \begin{pmatrix} \cos\left(\dfrac{\theta}{2}\right) & -\sin\left(\dfrac{\theta}{2}\right) \\ \sin\left(\dfrac{\theta}{2}\right) & \cos\left(\dfrac{\theta}{2}\right) \end{pmatrix} \left| \psi_1 \right\rangle$$

where |ψ₁> is created by a source, magnet and detector and |ψ₂> obtains when introducing a second SG measurement at an angle θ with respect to the first. The three possible orientations for SG measurments in $\Re^2$ considered here and in the Mermin apparatus (initial X+ orientation aside) are shown in figure 15. As with MZI outcomes, the description of spin measurement is to be understood via the spatiotemporal relationships between source(s) and detector(s) per the experimental arrangement, i.e., there are no "atoms impinging on the detectors" behind the SG magnets per their spins. There are just

sources, detectors and magnets whose relative orientations in space provide the computation of probabilities for event (click) distributions.

This constitutes an acausal and *non-dynamical* characterization and explanation of entanglement. According to our view, the *structure of correlations* evidenced by QLE is *determined by* the spacetime relations instantiated by the experiment, understood as a spatiotemporal whole. This determination is obtained by systematically *describing* the spatiotemporal symmetry structure of the Hamiltonian for the experimental arrangement[59]. Since

(i)      the explanation lies in the spacetime symmetries as evidenced, for example, in the family of trajectories per the Hamilton-Jacobi formalism,

(ii)     each family of trajectories characterizes the distribution of spacetime *relations,*

(iii)    we take those relations to be a timeless "block," and

(iv)    these relations collapse the matter-geometry dualism,

our geometrical quantum mechanics provides for an *a-causal*, *global* and *non-dynamical* understanding of quantum phenomena.

According to our geometrical view, the detector clicks are not caused by particles impinging on the detectors, from the source or otherwise. Using such a view, one can determine the correlations between the spin measurements in the quantum liar experiment, and thereby *explain* such correlations. This determination is obtained by systematically *describing* the spatiotemporal symmetry structure of the experimental arrangement.

*7.4 QLE and Blockworld.* Our analysis of QLE shows the explanatory necessity of the reality of all events—in this case the reality of all phases (past, present and future) of the QLE experiment. We can provide an illustrative, though qualitative, summary by dividing the QLE into three spatiotemporal phases, as depicted in figures 17 – 19. In the first phase the boxes Z1+, Z1-, Z2+, and Z2- are prepared – without such preparation the MZI is unaffected by their presence. In a sense, the boxes are being prepared as detectors since they have the potential to respond to the source ("atom absorbs the photon" in the language of dynamism). The second phase is to place the four boxes in the MZI per figure 14 and obtain a D1 or D2 click (null results are discarded). The third phase is to remove the four boxes and do spin measurements. The entire process is repeated many times with all possible Γ, Δ and Z spin measurements conducted randomly in phase 3. As a result, we note that correlations in the spin outcomes after D2 clicks violate Bell's inequality.

We are not describing "photons" moving through the MZI or "atoms" whose spin-states are being measured. According to our ontology, clicks are evidence not of an impinging particle-in-motion, but of a *spacetime relation*. If a Z measurement is made on either pair of boxes in phase 3, an inference can be made *a posteriori* as to which box acted as a "silent" detector in phase 2. If Γ and/or Δ measurements are done on each pair (figure 17), then there is *no fact of the matter* concerning the detector status of the original boxes (boxes had to be recombined to make Γ and/or Δ measurements). This is not simply a function of ignorance because if it was possible to identify the "silent" detectors before the Γ and/or Δ measurements were made, the Bell assumptions would be met and the resulting spin measurements would satisfy the Bell inequality. Therefore, *that none of the four boxes can be identified as a detector in phase 2 without a Z measurement in phase 3 is an ontological, not epistemological, fact*.

---

[59] The experimental apparatus itself providing the particular initial and final "boundary conditions" needed for a prediction unique to the apparatus.

Notice that what obtains in phase 3 "determines" what obtains in phase 2, so we have a true delayed-choice experiment. For example, suppose box Z2- is probed in phase 3 (Z measurement) and an event is registered (an "atom" resides therein," figure 18). Then, the Z2- and Z1- boxes are understood in phase 3 to be detectors in phase 2. However, nothing in the blockworld has "changed" – the beings in phase 2 have not "become aware" of which boxes are detectors. Neither has anything about the boxes in phase 2 "changed." According to our view, the various possible spatiotemporal distributions of events are each determined by NRQM *as a whole throughout space and time irrespective of space-like separation.*

To further illustrate the spatiotemporal nature of the correlations, suppose we make spin measurements after a D1 click. Figure 19 shows a spatiotemporal configuration of facts in phases 1, 2 and 3 consistent with a D1 click:

> Phase 1: No prep
> Phase 2: Boxes are not detectors, D1 click
> Phase 3: Γ2 measurement, Δ1 measurement, No outcomes.

One can find correlated spatiotemporal facts by starting in any of the three phases:

Starting with phase 3, "No outcomes" → "No prep" in phase 1 and "Boxes are not detectors" and "D1 click" in phase 2. If you insisted on talking dynamically, you could say that the "No outcomes" result of phase 3 determined "Boxes are not detectors" result of phase 2.

Starting with phase 2, "Boxes are not detectors" → "D1 click" in phase 2, "No prep" in phase 1 and "No outcomes" in phase 3.

Starting with phase 1, "No prep" → "No outcomes" in phase 3 and "Boxes are not detectors" and "D1 click" in phase 2.

One can chart implications from phase 1 to phase 3 then back to phase 2, since the order in which we chart implications in a spacetime diagram is meaningless (meta-temporal) to the blockworld inhabitants. In point of fact the three phases of QLE are jointly acausally and globally (without attention to any common cause principle) determined by the spacetime symmetries of all three phases of the experimental set-up. Hence, the explanatory necessity of the blockworld. What *determines* the outcomes in QLE is not given in terms of influences or causes. In this way we resolve the quantum liar paradox with geometrical quantum mechanics by showing how "the paradox" is not only *consistent* with a blockworld structure, but actually strongly suggests a non-dynamical approach such as ours over interpretations involving dynamical entities and their histories. Events in the context of this experiment are evidence not of particles, wavefunctions, etc., existing independently of the experimental set-up, but rather of spacetime relations (between source, detectors, etc.). It is the *spatiotemporal configuration of the QLE as a* *spacetime whole and its spacetime symmetries* that determine the outcomes and not *constructive entities with dynamical histories*.

## 8. Conclusion.

If, like us, you want to be a realist about:

1) the non-commutative structure of NRQM and the phenomena it entails such as entanglement and "non-locality"
2) the 4D spacetime structure as given by the Minkowski interpretation of STR and the blockworld it entails,

then there are some questions you must face. For example, what do we say about the multidimensional Hilbert space? Specifically, can one do justice to the non-commutative structure of NRQM without being a realist about Hilbert space? Our geometric interpretation constitutes an affirmative answer to this question. The trick is to appreciate that while everything "transpires" or rather resides in a 4D spacetime and nowhere else, nonetheless, some phenomena, namely quantum phenomena, cannot be modeled with worldlines if one is to do justice to its non-commutative structure. Thus while clicks in detectors are perfectly classical events, the clicks are not evidence of constructive quantum entities such as particles with worldlines, rather, the clicks are manifestations of spacetime relations between elements of the experimental configuration—distributions per the spacetime symmetries. Thus on our view while there is no "Dedukind cut" between the quantum and the classical as some versions of the Copenhagen interpretation would have it. After all, we can explain asymptotically the transition from the quantum to the classical in terms of density of "events." And there is also no "Einstein separability" between the system being measured and the system doing the measuring on our interpretation. Our view respects the causal structure of Minkowski spacetime in the sense that there are no faster than light "influences" or "productive" causes between space-like separated events as there are in Bohm for example. So our view is not non-local in any robustly dynamical sense. However our view does violate Einstein separability and it does have static "correlations" outside the lightcone as determined acausally and globally by the spacetime symmetries.

Such acausal global determination relations do not respect any common cause principle. This fact should not bother anyone who has truly transcended the idea that the dynamical or causal perspective is the most fundamental one. We are providing a model of an irreducibly relational blockworld, which is what realism about the quantum structure and the 4D spacetime structure yields once one accepts the implication therein of Hilbert space anti-realism. Having had this idea, others immediately come to mind. For example, rather than jump on the bandwagon of formulating constructive interpretations of NRQM (such as GRW or Bohm) and then seeking to make them relativistically covariant and invariant, why not abandon the centrality of the dynamical perspective altogether and seek out explanatory patterns and methods from a "God's eye point of view"—what Huw Price would call physics from the "Archimedean perspective." Hence, our acausal global determination relations. Those obsessed with constructive covariant interpretations of NRQM are just missing the point. And why not abandon the centrality of the constructive perspective and seek fundamental principle explanations based on the idea that constructive entities and their dynamical laws are not fundamental in the relational blockworld. Hence, our fundamental ontology of spatiotemporal relations. Our interpretation and our approach abandon the matter/geometry dualism that creates so many problems in physics. Though it is not easy to see, quantum theory and relativity theory are both trying to tell us the same thing from 'different angles' so to speak, the world is profoundly non-dual. The conjunction of the quantum structure with the spacetime structure, points to a non-duality that even the most reductionist schemas cannot yet fathom. We think these theories are trying to tell us that it is a 4D blockworld in which spatiotemporal relations are fundamental.

## APPENDIX: LORENTZ TRANFORMATIONS FOR SECTION 2

The speed of the boys relative to the girls is 0.6c, so

$$\sqrt{1 - \frac{v^2}{c^2}} = \sqrt{1 - \frac{0.36c^2}{c^2}} = 0.8$$

and

$$\gamma = \frac{1}{0.8} = 1.25 .$$

With T = t = 0 at X = x = 0, the girls' coordinates at the event labeled by the boys as t = 0, x = 1000km are given by the following Lorentz transformations

$$T = \gamma\left(t - \frac{vx}{c^2}\right) = 1.25\left(0 - \frac{0.6c*1000}{c^2}\right) = -0.0025s$$
$$X = \gamma(x - vt) = 1.25(1000 - 0.6c*0) = 1250km$$

where c = 300,000 km/s. And, the girls' coordinates at the boys' event t = 0.002, x = 1000km are

$$T = \gamma\left(t - \frac{vx}{c^2}\right) = 1.25\left(0.002 - \frac{0.6c*1000}{c^2}\right) = 0$$
$$X = \gamma(x - vt) = 1.25(1000 - 0.6c*0.002) = 800km$$

**Figure Captions**

Figure 1. Picture @ t = 0. Events 1 & 2 are simultaneous according to the boys.

Figure 2. Spacetime diagram showing time-like, null and space-like separated events in a BW spacetime. Events A, B and C are time-like separated from the origin, O. Event D is space-like separated from O. Event E is null separated from O. Events A and B are space-like separated from one another so some observers will see A occur before B while others will see B occur before A. In some frame of reference A and B are simultaneous, since they are space-like separated. The same is true for events O and D.

Figure 3. Spacetime diagram of presentism. Event C occurs at time $t_{-3}$ for all observers, regardless of their relative motions. Events O, A and B occur at time $t_0$ for all observers and are therefore unambiguously simultaneous. Event D occurs at time $t_4$ for all observers. Events C and D do not exist when events O, A and B exist since C is no longer 'real' and D is not yet 'real' when O, A and B are 'real'. 'Realness' only exists at one spatial surface $t_i$ at a time. [Clearly another 'time' is needed for this last statement to make sense.]

Figure 4. Picture @ T = -0.0025s. Event 2 occurred *before* Joe gets to Sara according to the girls.

Figure 5. Picture @ T = 0. Events 1 & 3 are simultaneous according to the girls.

Figure 6. Events 1 & 2 are simultaneous for the boys (both lie along t = 0 spatial plane). Boys' spatial planes of simultaneity are horizontal (t = 0 and t = 0.002s are shown). Girls' spatial planes of simultaneity (T = 0 and T = -0.0025s are shown) are tilted relative to the boys' spatial planes, as are the girls' worldlines tilted relative to the boys' worldlines. Events 1 & 3 are simultaneous for the girls (both lie along T = 0 spatial plane).
Events 1 & 2 are 'co-real' since Joe and Bob are both at these events at t = 0. Events 1 & 3 are 'co-real' since Sara and Alice are both at these events at T = 0. Bob is at both events, so his t = 0 self is 'co-real' with his t = 0.002s self.

**Figure 1**

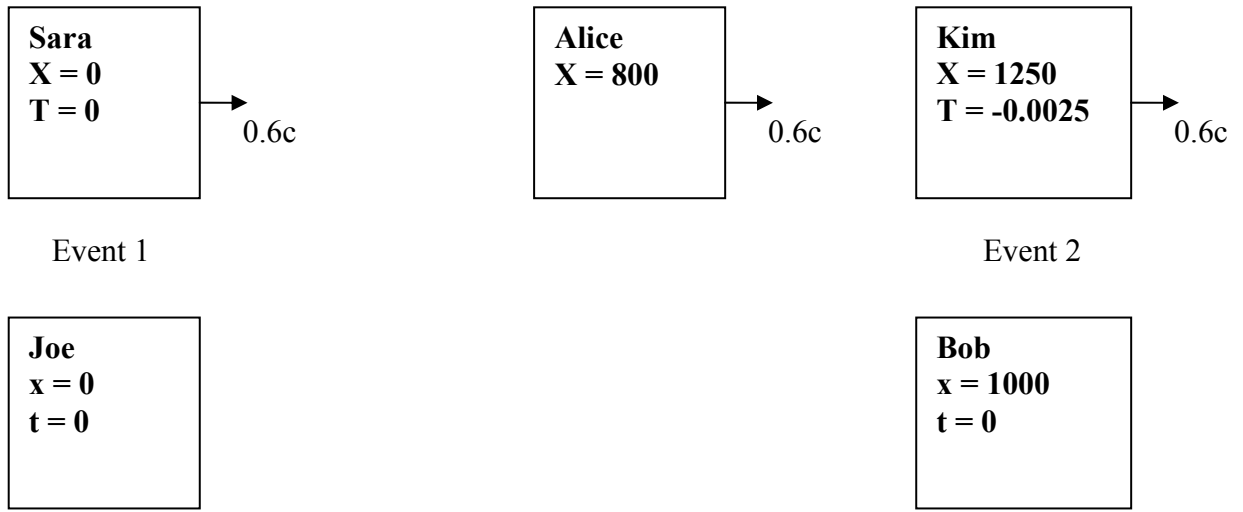

**Figure 2**

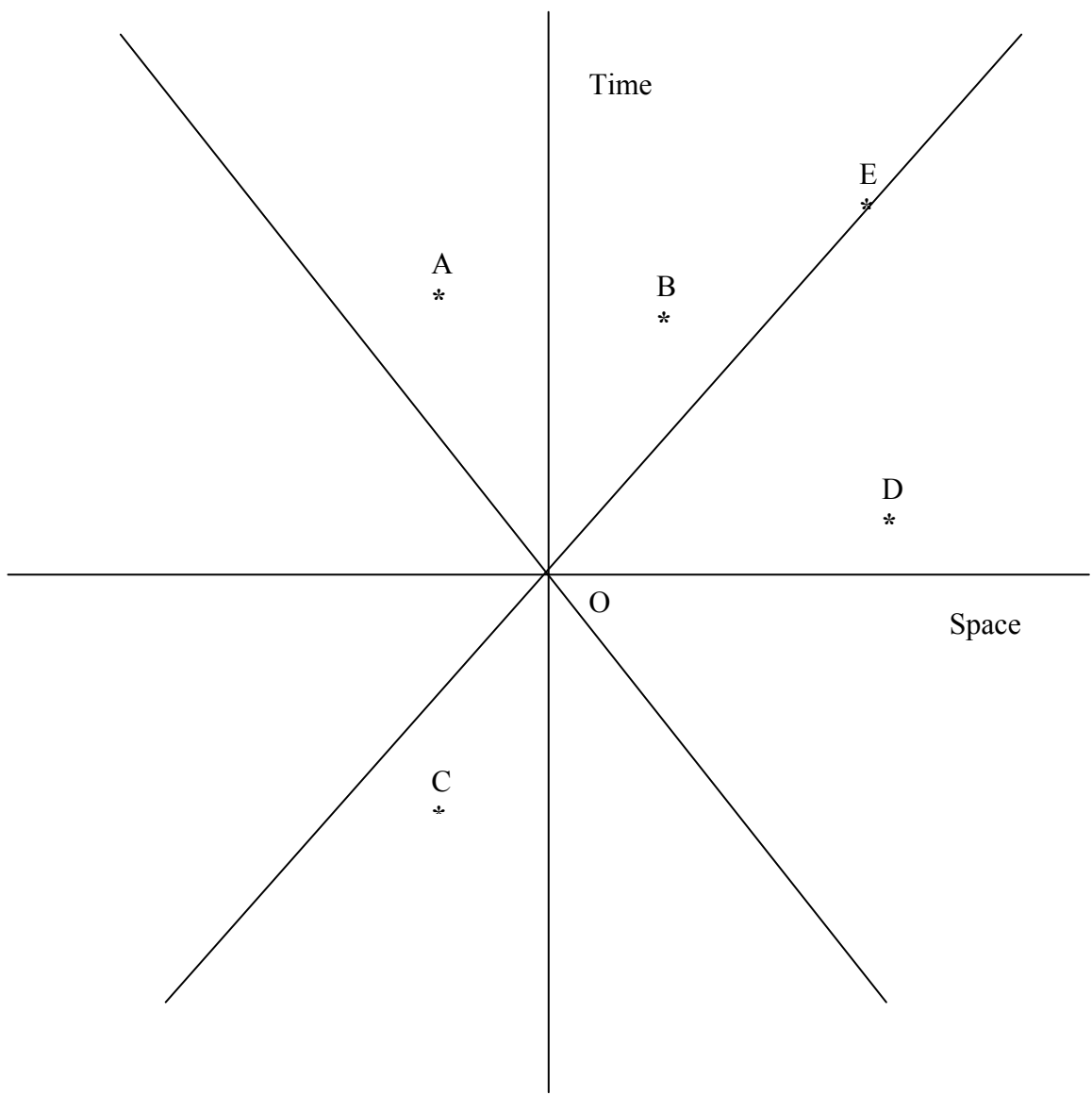

**Figure 3**

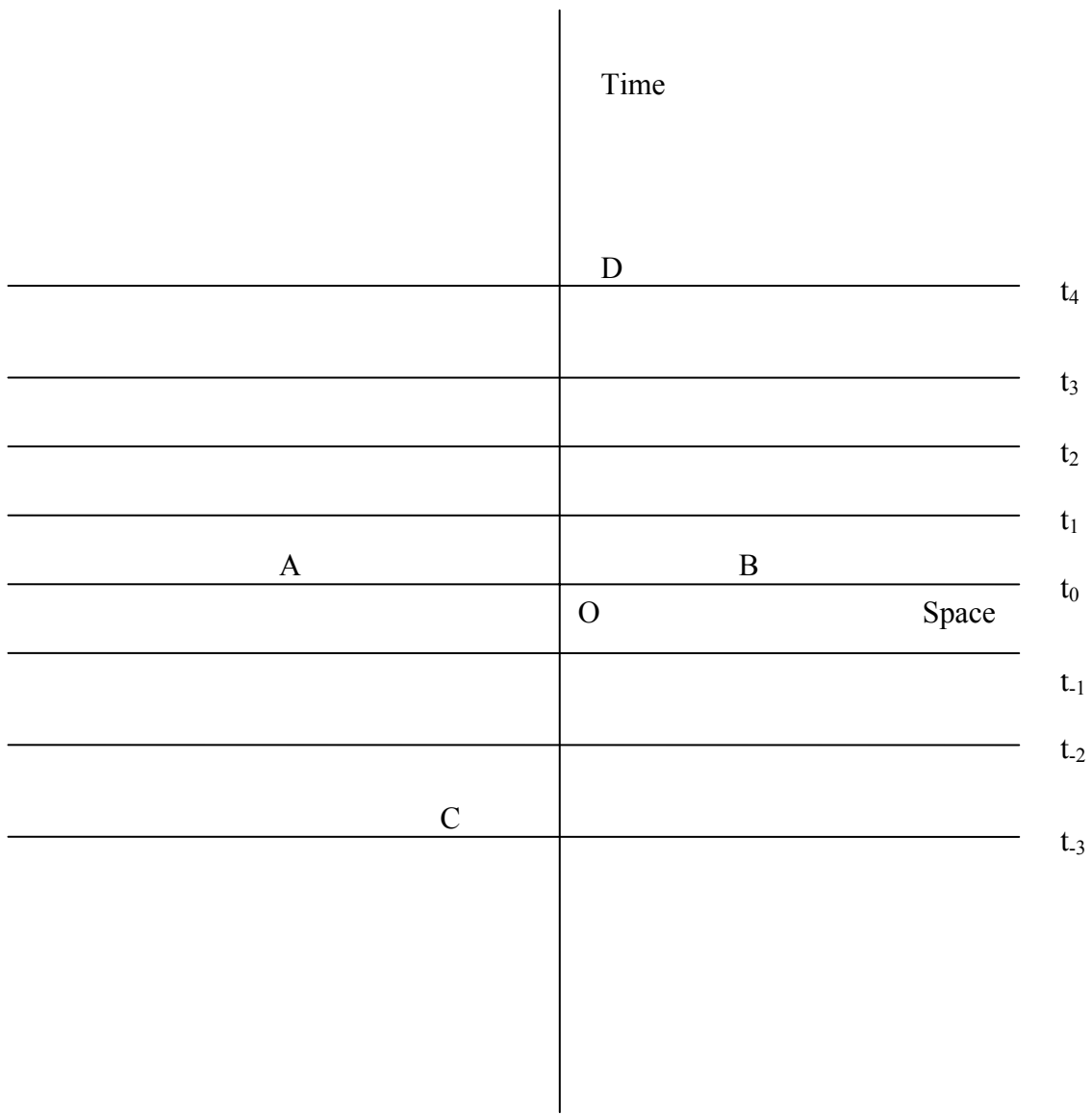

**Figure 4**

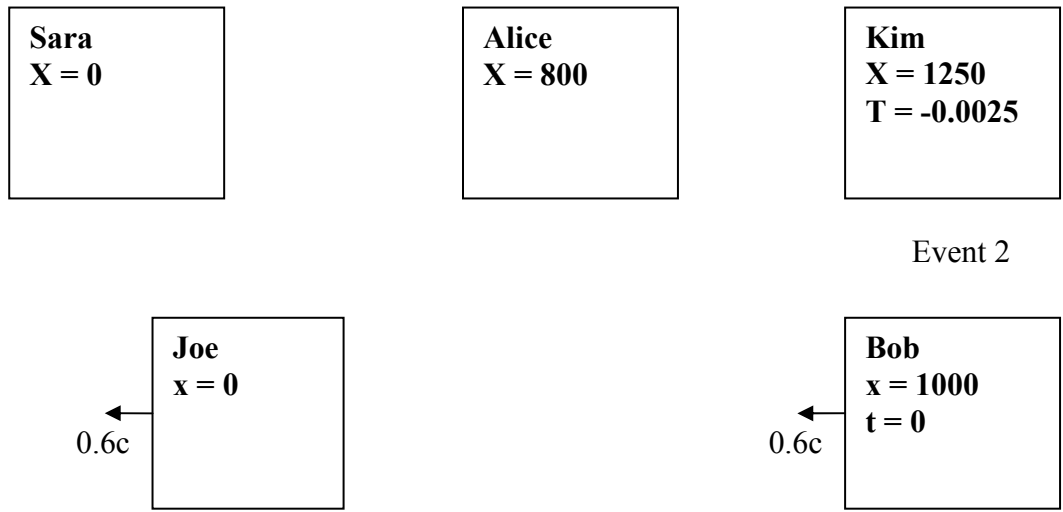

**Sara**
**X = 0**

**Alice**
**X = 800**

**Kim**
**X = 1250**
**T = -0.0025**

Event 2

**Joe**
**x = 0**

0.6c

**Bob**
**x = 1000**
**t = 0**

0.6c

**Figure 5**

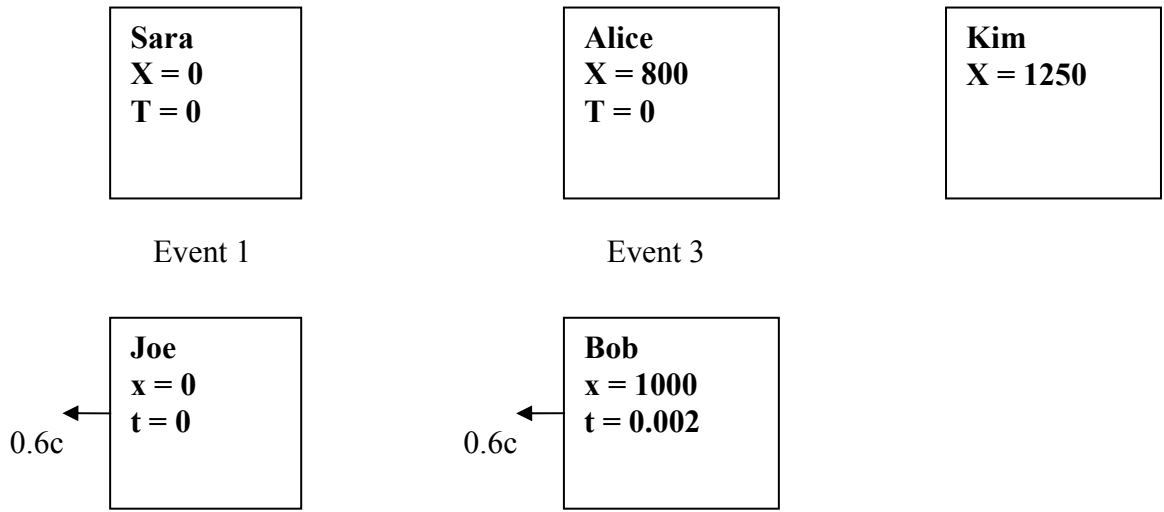

Sara
X = 0
T = 0

Event 1

Alice
X = 800
T = 0

Event 3

Kim
X = 1250

Joe
x = 0
t = 0

0.6c

Bob
x = 1000
t = 0.002

0.6c

**Figure 6**

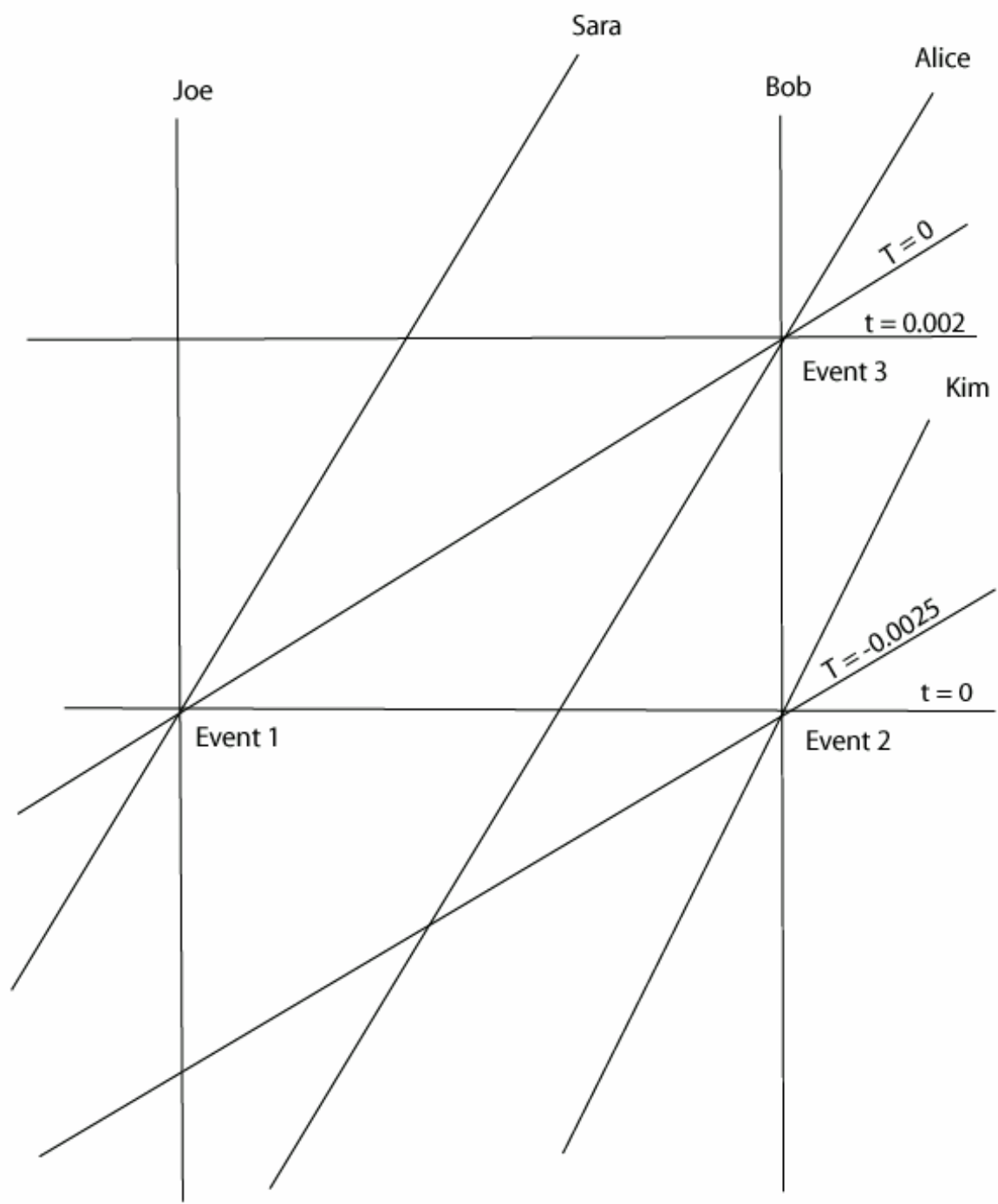

**Figure 7**

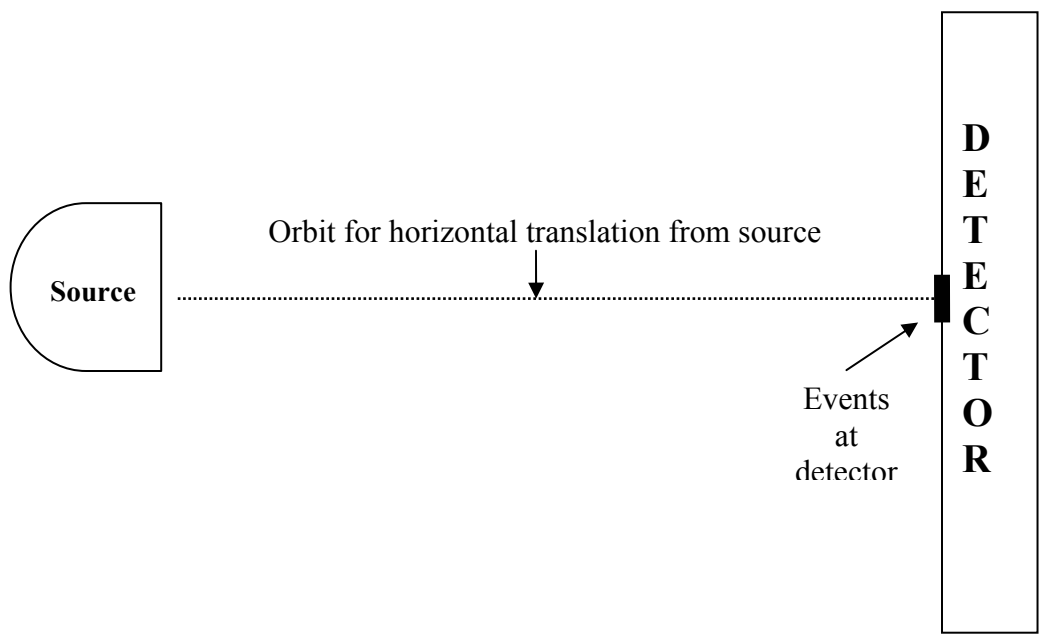

**Figure 8**

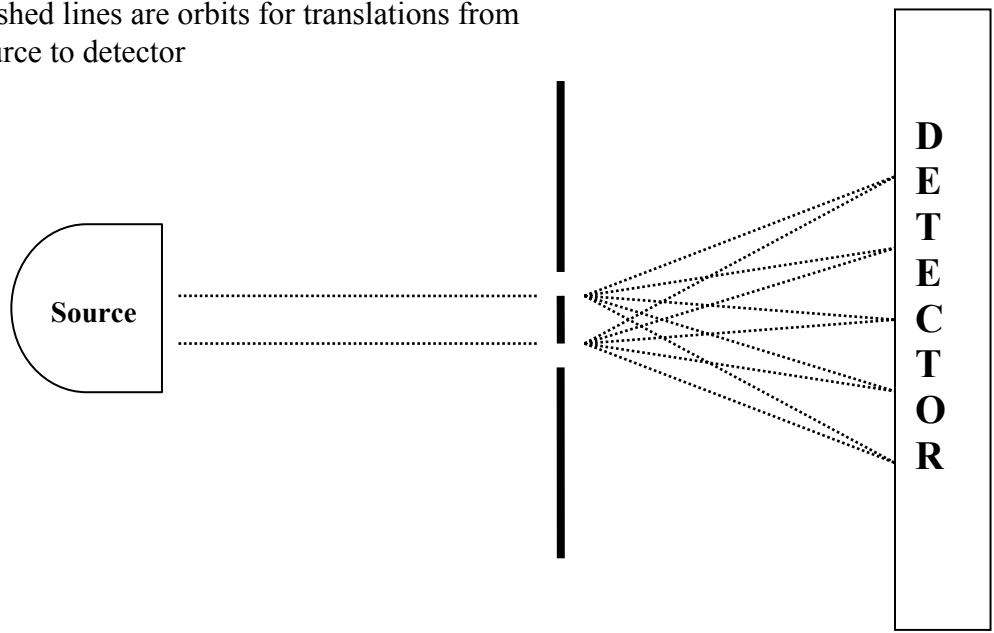

Dashed lines are orbits for translations from source to detector

**Figure 9**

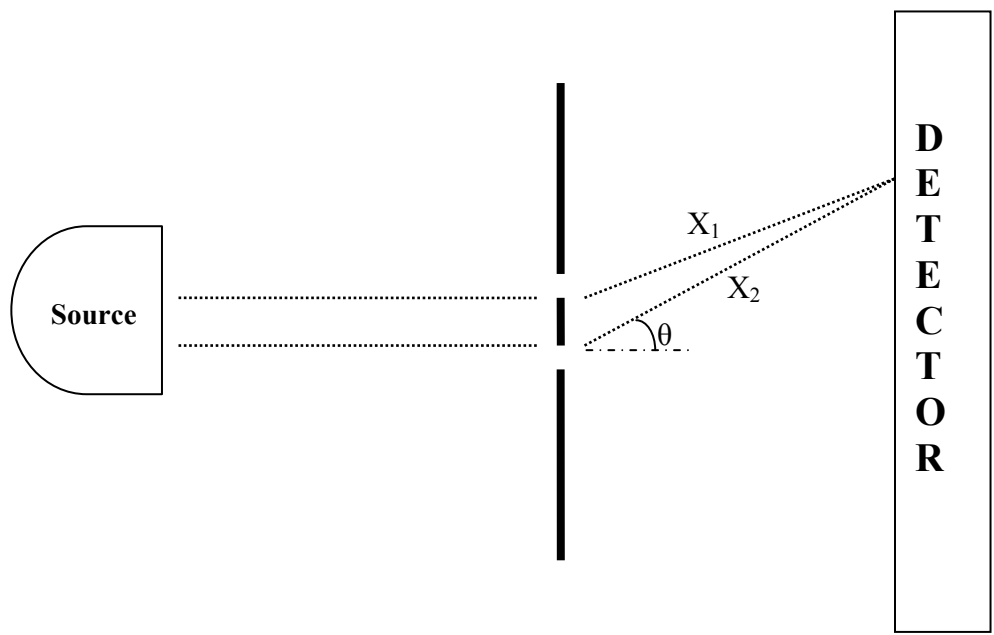

# Figure 10

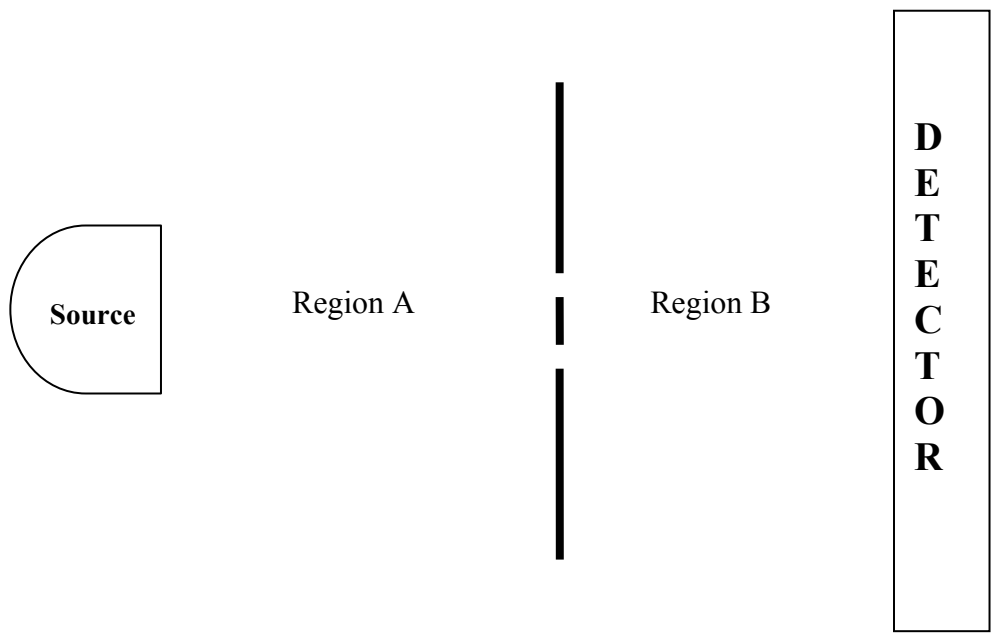

**Figure 11**

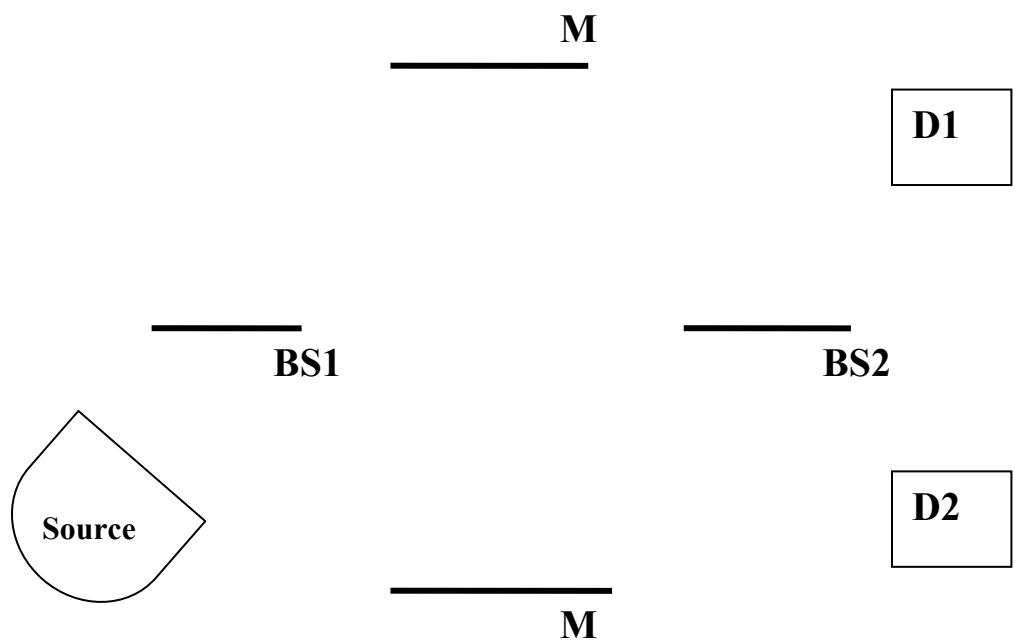

**Figure 12a**

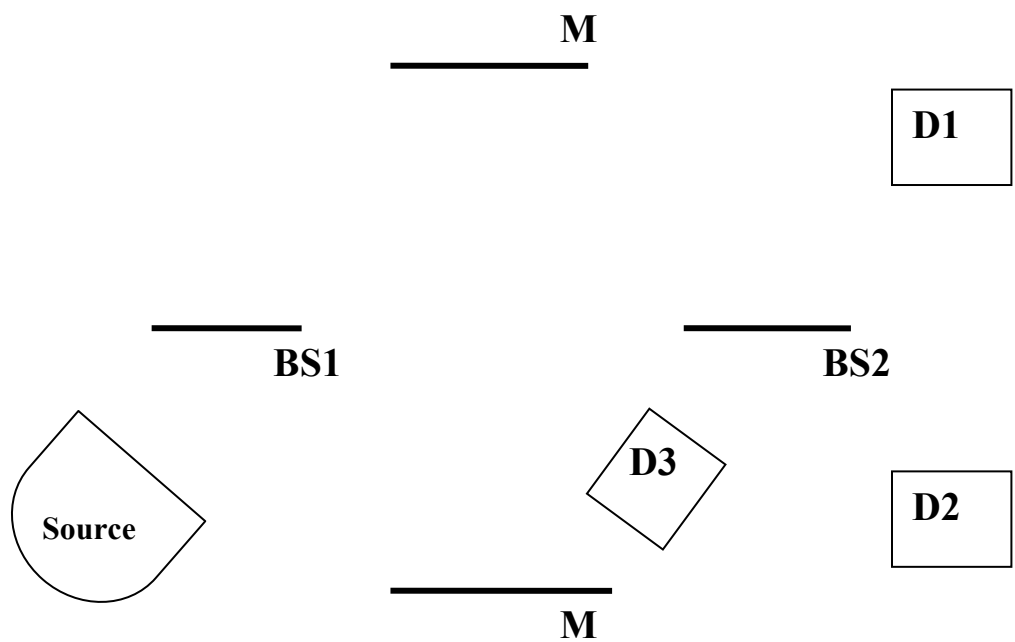

**Figure 12b**

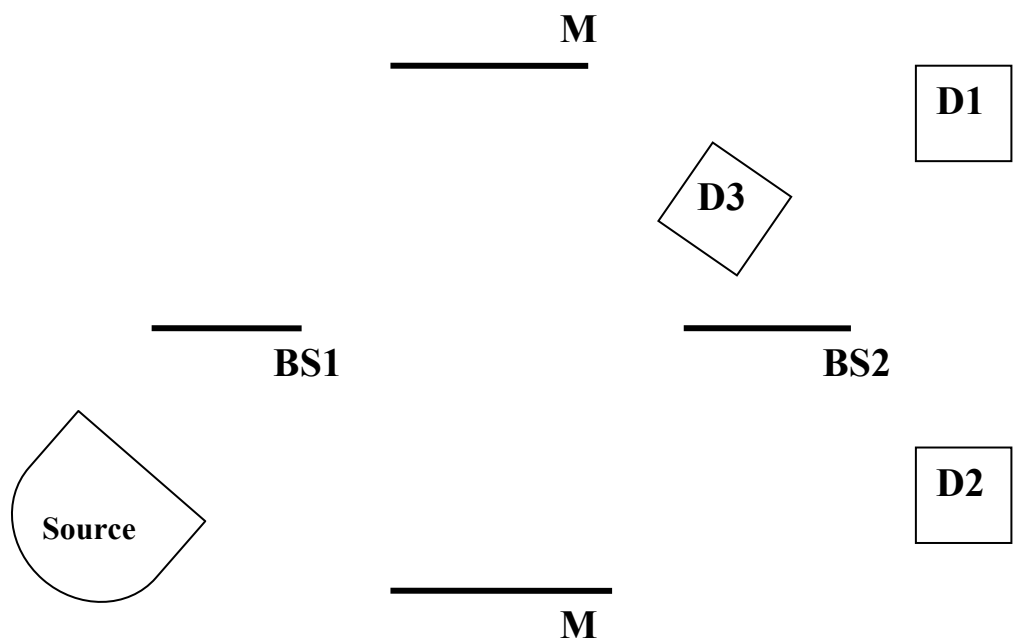

# Figure 13

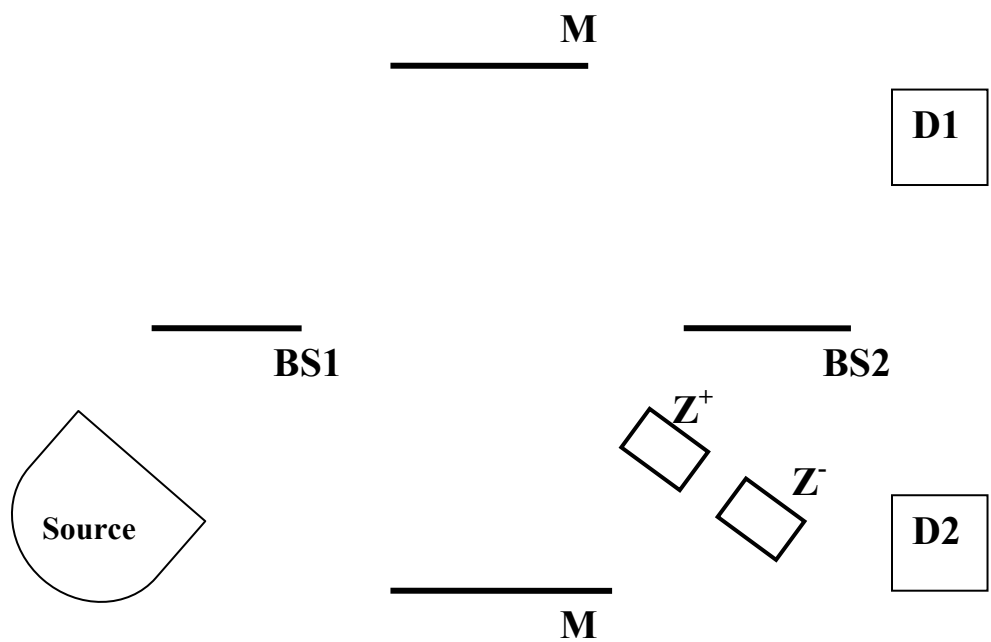

**Figure 14**

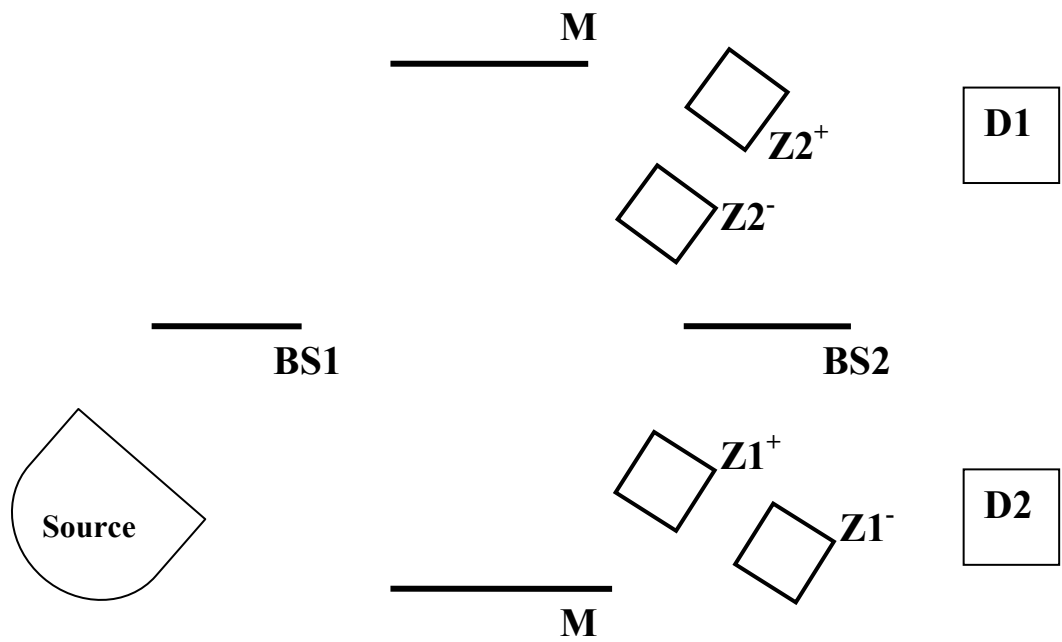

**Figure 15**

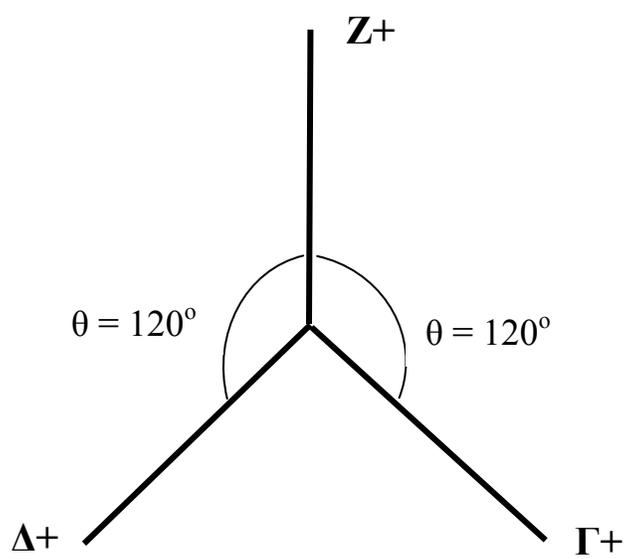

**Figure 16**

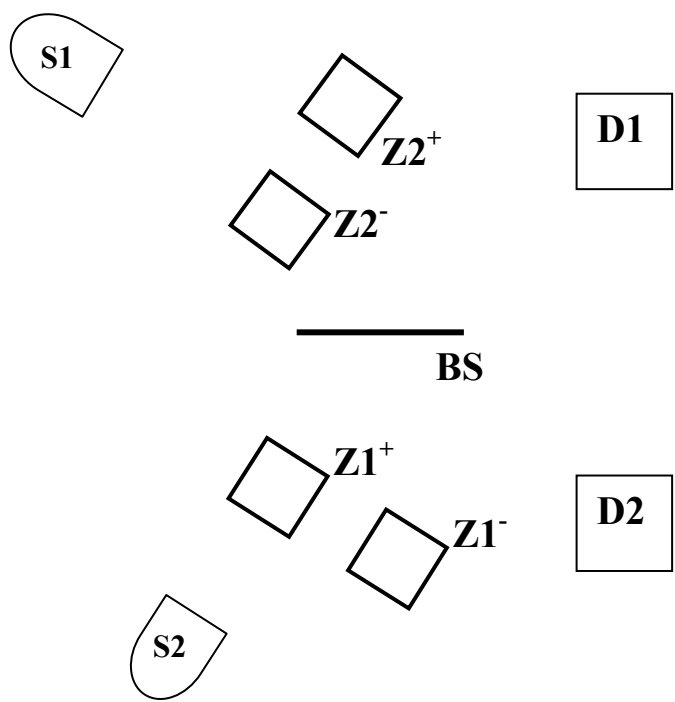

**Figure 17**

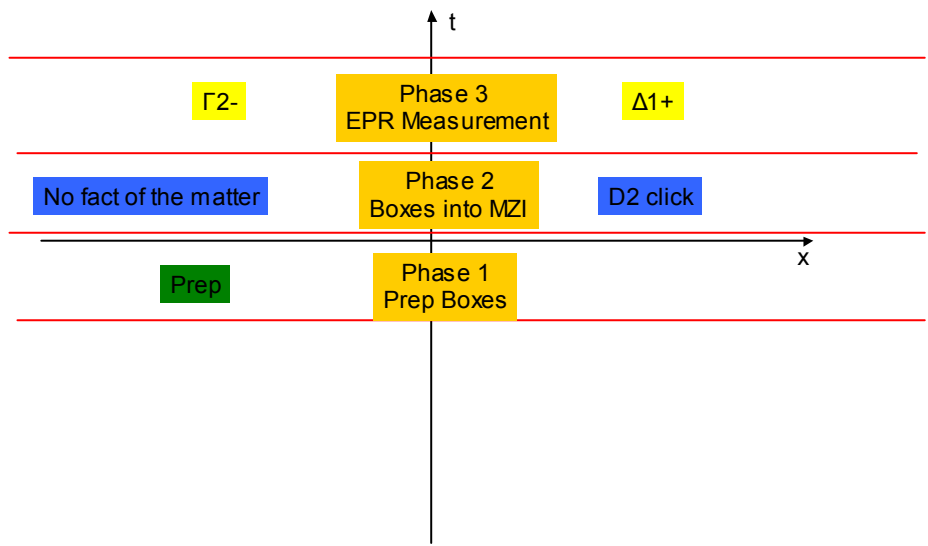

— 61

**Figure 18**

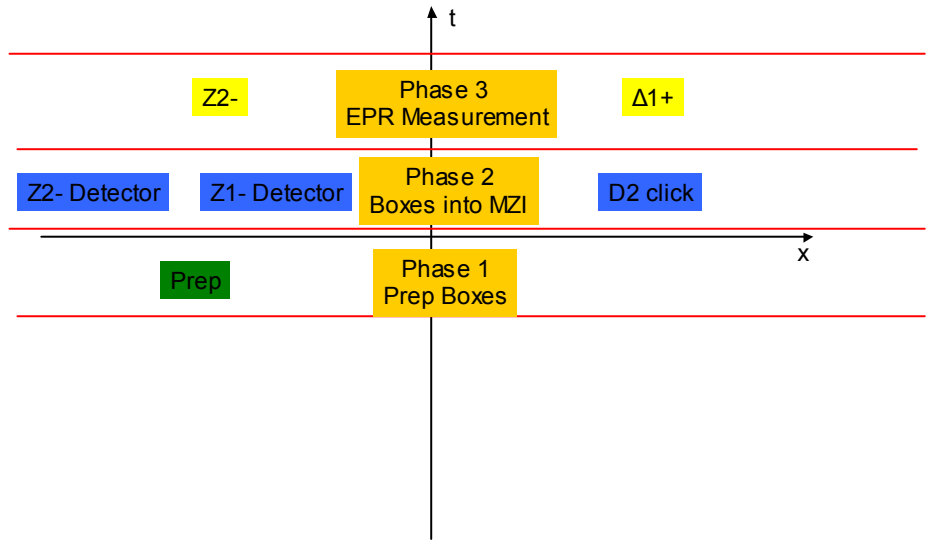

**Figure 19**

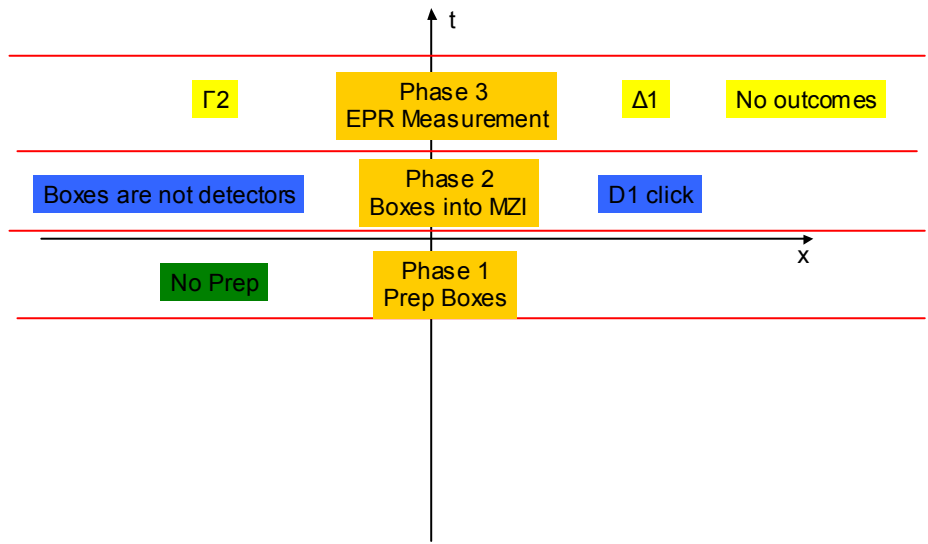

**REFERENCES**

Aharonov, Y. and Rohrlich, D., (2005), *Quantum Paradoxes: Quantum Theory for the Perplexed*. New York: John Wiley & Sons.

Anandan, Jeeva (1991), "A Geometric Approach to Quantum Mechanics," *Foundations of Physics* 21 (11): 1265 – 1284.

----------------- (1992), "The geometric phase," *Nature* 360 (6402): 307 – 313.

------------------ (2002*a*), "Symmetries, Quantum Geometry, and the Fundamental Interactions", *International Journal of Theoretical Physics* 41 (2): 199 – 220.

------------------ (2002*b*), "Causality, Symmetries and Quantum Mechanics," *Foundations of Physics* 15 (5): 415 – 438.

-------------- (2003), "Laws, Symmetries, and reality," *International Journal of Theoretical Physics* 42 (9): 1943 – 1955.

Ballentine, Leslie (1990), *Quantum Mechanics*. Englewood Cliffs, NJ: Prentice Hall.

Barrett, Jeffrey (1999), *The Quantum Mechanics of Minds and Worlds*. Oxford: Oxford. University Press.

Bohr, A. and Ulfbeck, O. (1995), "Primary manifestation of symmetry. Origin of quantum indeterminacy," *Reviews of Modern Physics* 67 (1): 1 – 35.

Bohr, A., Mottelson, B.R. and Ulfbeck, O. (2004*a*), "The Principle Underlying Quantum Mechanics," *Foundations of Physics* 34 (3): 405 – 417.

------------------------------------------------- (2004*b*), "Quantum World Is Only Smoke and Mirrors," *Physics Today* 57 (10): 15 – 16.

Brown, Harvey (2005), *Physical Relativity: Space-time Structure from a Dynamical Perspective*. Oxford: Oxford University Press.

------------------ and Holland, Peter R. (1999), "The Galilean covariance of quantum mechanics in the case of external fields," *American Journal of Physics* 67 (3): 204 – 214.

------------------ and Timpson, Chistopher G. (2006), "Why special relativity should not be a template for a fundamental reformulation of quantum mechanics," arXiv: quant-ph/0601182.

------------------ and Wallace, David (2005), "Solving the measurement problem: de Broglie-Bohm looses out to Everett," *Foundations of Physics* 35 (4): 517 – 540.

Bub, Jeffrey (2006), "Quantum Information and Computation," arXiv: quant-ph/0512125.

-------------- (2005), "Quantum mechanics is about quantum information," *Foundations of Physics* 35 (4): 541 – 560.

------------- (2004), "Why the quantum," *Studies in History and Philosophy of Modern Physics* 35B (2): 241 – 266.

Buccheri, R.; Elitzur, A.C.; and Saniga, M. (2005), *Endophysics, Time, Quantum and the Subjective*. Singapore: World Scientific.

DeWitt, R. (2004), *Worldviews: An Introduction to the History and Philosophy of Science*. London: Blackwell.

Dipert, Randall R. (1997), "World As Graph," *Journal of Philosophy* 94 (7), 329 – 358.

Elitzur, A. and Dolev, S. (2005*a*), "Quantum Phenomena Within a New Theory of Time" *in* Elitzur, *et. al*, pp. 325 – 349.

----------------------------- (2005*b*), "Becoming as a bridge between quantum mechanics and relativity" *in* Buccheri *et. al.*, *ibid*, pp. 589 – 606.

Elitzur, A.; Dolev, S. and Kolenda, N. (eds.) (2005), *Quo Vadis Quantum Mechanics*. Berlin: Springer.

Elitzur, A.C. and Vaidman, L. (1993), "Quantum-Mechanical Interaction-Free Measurements," *Foundations of Physics* 23 (7): 987 – 997.

Feynman, R. P., Leighton, R.B. and Sands, M., (1965) *The Feynman Lectures on Physics*, *Vol. III, Quantum Mechanics*, Reading: Addison-Wesley.

French, Steven and Ladyman, James (2003), "Remodelling Structural Realism: Quantum Physics and the Metaphysics of Structure," *Synthese* 136 (1): 31 – 56.

Frisch, Mathias (2005), "Mechanisms, principles, and Lorentz's cautious realism," *Studies in History and Philosophy of Modern Physics* 36B (4): 659 – 679.

Georgi, Howard (1999), *Lie Algebras in Particle Physics*, 2$^{nd}$ Ed. New York: Perseus Books.

Hardy, Lucien (1992), *Physics Letters A* 167: 11-16.

Holland, Peter R. (1993) *The Quantum Theory of Motion: An Account of the de Broglie-Bohm Causal Interpretation of Quantum Mechanics*. Cambridge: Cambridge University.

Janssen, Michel (2002*b*), "Reconsidering a Scientific Revolution: The Case of Einstein *versus* Lorentz," *Physics in Perspective* 4: 421 – 446.

Kaiser, Gerald (1981), "Phase-space approach to relativistic quantum mechanics. III. Quantization, relativity, localization and gauge freedom," *Journal of Mathematical Physics* 22 (4): 705 – 714.

---------------- (1990), *Quantum Mechanics, Relativity, and Complex Spacetime: Towards a New Synthesis*. Amsterdam: North-Holland Press.

Lepore, J.V. (1960), "Commutation relations of quantum mechanics," *Physical Review* 119 (2): 821 – 826.

Lewis, Peter J. (2004), "Life in configuration space," *British Journal for the Philosophy of Science* 55 (4): 713 – 729.

Maudlin, Tim (2005), "Non-local Correlations: How the Trick Might Be Done" (unpublished manuscript).

----------------- (1994), *Quantum Non-Locality & Relativity*. Oxford: Blackwell.

Mermin, N. D. (1981), "Bringing Home The Atomic World: Quantum Mysteries for Anybody," *American Journal of Physics* 49 (10): 940 – 943.

Park, D. (1992), *Introduction to Quantum Theory*. 3rd edition. McGraw-Hill.

----------------- (1968), *Space and Time in Special Relativity*. New York: McGraw-Hill.

Petkov, V. (2006), "Is There an Alternative to the Blockworld View?" *in* Dieks, D. (ed.) *The Ontology of Spacetime*. Amsterdam: Elsevier.

------------ (2005), *Relativity and the Nature of Spacetime*. Berlin: Springer.

Stuckey, W.M., Silberstein, Michael, and Cifone, Michael (2006), "Deflating Quantum Mysteries via the Relational Blockworld," to appear in *Physics Essays* 19, arXiv: quant-ph/0503065.

Tooley, Michael (1997), *Time, Tense and Causation*. New York: Clarendon Press.

Zee, A. (2003), *Quantum Field Theory in a Nutshell*. Princeton: Princeton University Press.